\documentclass[fleqn,usenatbib,useAMS]{mnras}
\usepackage{newtxtext,newtxmath}

\usepackage{graphicx}	
\usepackage{amsmath}	
\usepackage{multicol}        
\usepackage{bm}		
\usepackage{pdflscape}	
\usepackage{float}
\usepackage[normalem]{ulem} 
\usepackage[dvipsnames]{xcolor}
\usepackage{soul}

\usepackage{tikz}
\usetikzlibrary{calc, positioning}

\usepackage[percent]{overpic}

\usepackage{booktabs, multirow, makecell}

\setcellgapes{5pt}
\newcommand{\msol}{\, {\rm M}_{\odot}}
\newcommand{\lsol}{\, {\rm L}_{\odot}}
\newcommand{\kms}{\,{\rm km \, s^{-1}}}
\newcommand{\pc}{\,{\rm pc}}
\newcommand{\kpc}{\,{\rm kpc}}
\newcommand{\kev}{\,{\rm keV}}

\newcommand{\oversim}[2]{\protect{\mbox{\lower0.5ex\vbox{%
   \baselineskip=0pt\lineskip=0.2ex
   \ialign{$\mathsurround=0pt #1\hfil##\hfil$\crcr#2\crcr\sim\crcr}}}}} 
\newcommand{\bb}[1]{\ifmmode \mbox{\boldmath $ #1$} \else  \mbox{\boldmath $#1$} \fi}
\catcode`\"=\active\let"=\"  

\def\3{{\ss} }

\def\c12{{1\over 2}}

\def\d{{\rm d}}   
   
\def\plusplus{\raise 0.3ex\hbox{${\scriptstyle ++}$}{}}

\def\and{{{\rm M}31}}

\def\gyr{\,{\rm Gyr}}
\def\myr{\,{\rm Myr}}

\bibliography{biblio}

\usepackage{array}
\newcolumntype{L}[1]{>{\raggedright\let\newline\\\arraybackslash\hspace{0pt}}m{#1}}
\newcolumntype{C}[1]{>{\centering\let\newline\\\arraybackslash\hspace{0pt}}m{#1}}
\newcolumntype{R}[1]{>{\raggedleft\let\newline\\\arraybackslash\hspace{0pt}}m{#1}}

\usepackage[normalem]{ulem}

\begin{document}   
\title[Capture of stars by dark substructures]{Capture of field stars by dark substructures}

\author[Pe\~narrubia et al.]{Jorge Pe\~narrubia$^{1,2}$\thanks{Email: jorpega@roe.ac.uk}, Rapha\"el Errani$^{3,4}$, Matthew G. Walker$^{3}$, Mark Gieles$^{5,6}$, Tjarda C. N. Boekholt$^{7,8}$\\
$^1$Institute for Astronomy, University of Edinburgh, Royal Observatory, Blackford Hill, Edinburgh EH9 3HJ, UK\\
$^2$Centre for Statistics, University of Edinburgh, School of Mathematics, Edinburgh EH9 3FD, UK\\
$^3$McWilliams Center for Cosmology, Department of Physics, Carnegie Mellon University, 5000 Forbes Ave., Pittsburgh, PA 15213, United States\\
$^4$Universit\'e de Strasbourg, CNRS, Observatoire Astronomique de Strasbourg, UMR 7550, F-67000 Strasbourg, France\\
$^5$ICREA, Pg. Lluís Companys 23, E08010 Barcelona, Spain\\
$^6$Institut de Ci\`encies del Cosmos (ICCUB), Universitat de Barcelona (IEEC-UB), Mart\'i i Franqu\`es 1, E08028 Barcelona, Spain\\
$^7$Rudolf Peierls Centre for Theoretical Physics, Clarendon Laboratory, University of Oxford, Parks Road, Oxford, OX1 3PU, UK\\
$^{8}$NASA Ames Research Center, Moffett Field, CA 94035, USA
}
\maketitle

\begin{abstract}
We use analytical and $N$-body methods to study the capture of field stars by gravitating substructures moving across a galactic environment. 
The majority of stars captured by a substructure move on temporarily-bound orbits that are lost to galactic tides after a few orbital revolutions.
In numerical experiments where a substructure model is immersed into a sea of field particles on a circular orbit, we find a population of particles that remain bound to the substructure potential for indefinitely-long times.
This population is absent from substructure models initially placed outside the galaxy on an eccentric orbit.
We show that gravitational capture is most efficient in dwarf spheroidal galaxies (dSphs) on account of their low velocity dispersions and high stellar phase-space densities.
In these galaxies `dark' sub-subhaloes which do not experience in-situ star formation may capture field stars
and become visible as stellar overdensities with unusual properties: (i) they would have a large size for their luminosity, (ii) contain stellar populations indistinguishable from the host galaxy, and (iii) exhibit dark matter (DM)-dominated mass-to-light ratios.
We discuss the nature of several `anomalous' stellar systems reported as star clusters in the Fornax and Eridanus II dSphs which exhibit some of these characteristics. 
DM sub-subhaloes with a mass function $\d N/\d M_\bullet\sim M_\bullet^{-\alpha}$ are expected to generate stellar systems with a luminosity function, $\d N/\d M_\star\sim M_\star^{-\beta}$, where $\beta=(2\alpha+1)/3=1.6$ for $\alpha=1.9$.
Detecting and characterizing these
objects in dSphs would provide unprecedented constraints on the particle mass
and cross section of a large range of DM particle candidates.
\end{abstract}


\section{Introduction}\label{sec:intro}
One of the strongest predictions from Cold Dark Matter (CDM) cosmology is the existence of self-gravitating haloes devoid of visible matter (i.e., `dark'). Such objects arise because star formation becomes inefficient in haloes with virial masses below $\sim 10^7$--$10^9\,\mathrm{M}_\odot$ (White \& Rees 1978; Bullock et al. 2000; Bovill \& Ricotti 2009; Benitez-Llambay \& Frenk 2020, Pereira-Wilson et al. 2023), while the minimum subhalo mass associated with the free-streaming length of `cold' particles with masses above $\sim 1\,\mathrm{GeV/c}^2$ is at least $\sim 13$ orders of magnitude smaller, $<10^{-6}\, \rm M_\odot$ (e.g. Schmid et al. 1999; Hofmann et al. 2001; Green et al. 2005; Loeb \& Zaldarriaga 2005; Diemand et al. 2005). Below this scale, fluctuations of the power spectrum are heavily suppressed, de facto imposing a truncation at the low end of the halo mass function (e.g. Benson 2017 and references therein).
Hence, detecting this truncation would provide a direct constraint on the DM particle mass.

The lack of visible matter together with their tiny masses make the detection of dark subhaloes extremely difficult. Current observational efforts range from searching for gamma- ray annihilation signals (e.g. Ackermann et al. 2014; Bringmann et al. 2014) to modelling substructure in strongly-lensed galaxies (Koopmans 2005; Vegetti \& Koopmans 2009; Li et al. 2013; Vegetti et al. 2014). In the Milky Way, encounters with individual subhaloes can induce significant perturbations in cold tidal streams (Ibata et al. 2002; Johnston et al. 2002; Yoon et al. 2011; Carlberg 2013; Ngan et al. 2016, Erkal et al. 2016). For example, Bonaca et al. (2019) find that some observed features in the GD-1 stream, including a gap and an off-stream spur of stars, are best reproduced by the past encounter with a dark subhalo with a mass $M_\bullet\sim 10^6$--$10^8\msol$ and a scale radius $c_\bullet\lesssim 10\pc$. Puzzlingly, these constraints imply a matter density comparable to the stellar density in globular clusters, $\rho_\bullet=M_\bullet/(2\pi c_\bullet^3)\gtrsim 10^2\msol\pc^{-3}$, which is several orders of magnitude denser than CDM subhaloes with similar masses found in cosmological simulations of structure formation (e.g. Molin\'e et al. 2017; Diemer \& Joyce 2019).

Inferring the existence of dark subhaloes using dynamical probes is complicated by the unknown number of baryonic compact objects -- such as stellar black holes, neutron stars, white dwarfs, free-floating planets, giant molecular clouds etc. -- lurking in the Galaxy. In general, it is not straightforward to isolate perturbations arising from different populations of gravitating objects (Pe\~narrubia 2018).
In this regard, dwarf spheroidal galaxies (dSphs) provide relatively clean targets, as their gravitational potentials appear to be fully dominated by dark matter (e.g. Mateo 1998).

This paper challenges the common assumption that DM subhaloes that do not form stars {\it in situ} remain `dark'. 
Here, we show that dark subhaloes can capture\footnote{In this paper, the word ‘capture’ is broadly used to describe any dynamical process wherein a field particles undergoes a transition from galaxy orbit to an orbit around a substructure.} baryonic matter as they orbit around the host galaxy, becoming `visible' as localized substructures of co-moving bodies with high mass-to-light ratios and extended sizes. 
Dark subhaloes that are massive enough to capture field stars would bear resemblance to stellar clusters, but with atypical properties, e.g. their stellar populations would be chemically and chronologically identical to the local galactic field, and they would be DM dominated. As we will show below, these properties are akin to those of the `anomalous' clusters detected in some Milky Way dSphs, suggesting the intriguing possibility that these systems may be in fact agglomerates of field stars captured by dark substructures orbiting in the host galaxies. This possibility is explored below in some detail.

Capture processes in the classical\footnote{Two point masses on circular orbits, plus a mass-less tracer} restricted 3-body problem have been studied for a long time. For example, the pioneering work of Szebehely (1967) showed that a finite number of solutions exists where the lightest particle is transferred from one distinct mode of motion around the most massive point-mass to another distinct mode around the intermediate-mass one. Hunter (1967), Heppenheimer (1975) and Heppenheimer \& Porco (1976) pointed out that Jupiter's outer satellites could have been captured in this way. More recently, Suetsugu \& Ohtsuki (2013) study temporary capture of planetesimals by a giant planet, while J{\'\i}lkov{\'a} et al. (2015) consider the scenario where the inner Oort Cloud was captured from another star during a close encounter in their birth cluster. Recently, 3-body captures in accretion discs have also gained attention as a possible source of black-hole binaries. E.g. Li et a. (2022), Boekholt et al. (2023) and Rowan et al. (2023) show that close encounters between two black holes orbiting around a supermassive black hole can form bound pairs.

The dynamics of 3-body capture events is extremely complex. Using a combination of numerical and analytical methods, Petit \& H\'{e}non (1986) showed that (i) captures in a 3-body system are temporary events that ultimately induce the dissolution of the bound pair, and (ii) capture only happens for extremely fined-tuned combinations of impact parameters and relative velocities which exhibit a self-similar, Cantor-like structure. More recent work of Boekholt et al. (2023) confirmed these results and found that the phase space structure that leads to capture resembles a Cantor set with a fractal dimension of $\simeq 0.4$.

Pe\~narrubia (2023, hereafter Paper I) studies 3-body captures in a galactic environment, where the intermediate and lighter bodies (both point masses) follow orbits in a massive (extended) potential. Using numerical experiments, Paper I shows that encounters between a massive object and field particles can be locally described as slingshot manoeuvres, in which the lighter body can increase/decrease its speed or redirect its path. For a capture event to happen, the galactic tidal field must decelerate an approaching lighter body to a degree where it temporarily orbits the intermediate one. 
Crucially, a point-mass moving through a sea of lighter particles generates a localized overdensity -- or `halo' -- of tidally-trapped particles\footnote{A massive perturber also deflects stars into an overdense `wake' that trails behind it (e.g. Kalnajs 1971; Weinberg 1986). In contrast, this work focuses on field stars that become temporarily bound to the perturber.}
, which reaches a steady state as the rate of bodies captured from the field becomes comparable to those being lost to galactic tides. 

The reverse process by which stars escape from the intermediate potential in a restricted 3-body system is not fully understood. For example, Fukushige \& Heggie (2000) show that stars with energies above the energy of escape can remain inside the tidal radius of the intermediate body for very long times, and some do not escape at all. These particles are typically dubbed `potential escapers' (see e.g. K\"upper et al. 2010; Daniel et al. 2017a).

In this work we expand the analysis of Paper I to intermediate bodies with an extended (i.e., not point-like) mass distribution. As an application, we study capture of field stars by DM subhaloes in a wide range of galactic environments.
The paper is arranged as follows. \S\ref{sec:model} extends the statistical method outlined in Paper I to model gravitational captures by point-mass objects to substructures an extended size.
\S\ref{sec:experiment} presents numerical experiments that test the accuracy of the theoretical equations. 
\S\ref{sec:dis} discusses these results in the context of dark matter particle physics and outlines future follow-up work. Finally, in \S\ref{sec:sum}, we summarize our main results.

\section{Statistical theory}\label{sec:model}
This Section summarizes the main techniques applied in this paper to construct a statistical theory that describes the spatial and kinematical distribution of tracer particles temporarily bound to self-gravitating substructures moving in a host galaxy. 

Section~\ref{sec:number} follows the steps outlined in Paper I to compute the  average number of field particles that have negative binding energies $E=v^2/2+\Phi_\bullet(r)<0$ within a spherical volume $V=4\pi r^3/3$, where $r$ and $v$ are measured relative to the substructure, as well as their number density ($n_\star$) and velocity dispersion ($\sigma_\star$) profiles in dynamical equilibrium within a generic potential $\Phi_\bullet$. Here, we will focus on capture of field {\it stars}, although this analysis can be extended to other tracers moving in the host galaxy potential, including gas particles\footnote{To model the dynamics of gaseous particles, a pressure term must be included in the equations of motion presented in \S\ref{sec:nbody}}. 

To gain further physical insight, we will inspect the case of dark substructures sourcing a Hernquist (1990) potential 
\begin{align}\label{eq:hern}
  \Phi_\bullet(r)=-\frac{GM_\bullet}{r+c_\bullet},
\end{align}
which recovers the results derived in Paper I for point-masses if the profile scale-length is set to $c_\bullet= 0$. Note that the density profile associated with the potential~(\ref{eq:hern}) has a centrally-divergent cusp, $\rho\sim r^{-1}$ at $r\ll c_\bullet$, as the universal profile found in CDM simulations of structure formation (e.g. Navarro, Frenk \& White 1997).

Section~\ref{sec:MW} applies our results to Milky Way dwarf spheroidal galaxies, which, due to their intrinsic properties and relatively close distances, appear to be the most interesting objects to test the existence of dark substructures via captured field stars.

Section~\ref{sec:sub} analyzes the cosmologically-motivated case of a large population of extended substructures with a power-law mass function.

\subsection{Distribution of trapped particles }\label{sec:number}
\subsubsection{Weak perturbations}
The number of field stars temporarily bound to a moving substructure sourcing a potential $\Phi_\bullet$ can be estimated statistically from the (local) distribution function under the following simplifying assumptions: (i) field stars move on uncorrelated (random) trajectories within a small volume element $V$, (ii) their number density is roughly constant within the volume $V$, such that $n({\mathbfit R}_\bullet+{\mathbfit r})\approx n({\mathbfit R}_\bullet)=n=N/V$, where ${\mathbfit R}_\bullet$ is the 3D position vector of the substructure within the host galaxy, and $r=|\bb R-\bb R_\bullet|$ is the relative distance of a star from the centre of the potential $\Phi_\bullet$. This is known as the {\it local approximation}, and holds insofar as the (local) density profile rolls slowly, i.e. $r\ll d=|\nabla n/n|^{-1}$. (iii) The relative velocity distribution of particles within the volume element $V=4\pi r^3/3$ follows a Maxwellian distribution displaced by the reflex velocity of the point-mass 
$p(\bb v)=(2\pi \sigma^2)^{-3/2}\exp[-(\bb v+\bb V_\bullet)^2/(2\sigma^2)$
 where $\sigma=\sigma(\bb R_\bullet)$ is the local, one-dimensional velocity dispersion of field stars. Under the Maxwellian approximation, the mean-squared (relative) velocity between the background particles and the substructure is $\langle v^2\rangle =3\sigma^2 + V_\bullet^2$,
where ${\mathbfit V}_\bullet$ is the velocity vector of the substructure with respect to the host galaxy centre.

Under these conditions, the average number of stars in the volume $V$ with negative specific energy $E=v^2/2+\Phi_\bullet<0$ can be derived from the local phase space density of field stars as
\begin{align}\label{eq:Nb}
N_b(r)&=\int_V\d^3 r\, n(\bb r) \int_{E<0}\d^3v\,p(\bb v) \\\nonumber
&= \frac{1}{3}\sqrt{\frac{2}{\pi}}z e^{-V_\bullet^2/(2\sigma^2)} \frac{n}{\sigma^3} \int_V\d^3 r\,v_e^3(\bb r ) +\mathcal{O}(v_e/\sigma)^5& \\ \nonumber
&=\frac{32\sqrt{\pi}}{9}(G M_\bullet)^{3/2}\,e^{-V_\bullet^2/(2\sigma^2)}\frac{n}{\sigma^3} \\ \nonumber 
  &~~~\times \frac{r^2-4 c_\bullet r+8c_\bullet^{3/2}(r+c_\bullet)^{1/2}-8 c_\bullet^2}{(r+c_\bullet)^{1/2}},
\end{align} 
for $r\ll d$ and $v_e/\sigma\lesssim 1$. 
To gain physical intuition, the last expression adopts a Hernquist potential~(\ref{eq:hern}) with a escape speed $v_e(r)=\sqrt{2|\Phi_\bullet|}=\sqrt{2 GM_\bullet/(r+c_\bullet)}$. It is trivial to show Equation~(\ref{eq:Nb}) recovers Equation~(11) of Paper I for a point-mass with $c_\bullet=0$.
Note that the number of stars with negative energies within the volume $V$ is proportional to the local {\it mean phase-space density} of the field, $Q\equiv n/\sigma^3$, and that $N_b$ drops for substructures that are not at rest with the background  ($V_\bullet> 0$).

The approximation that field stars move on uncorrelated trajectories is accurate insofar as the number of stars perturbed by $\Phi_\bullet$ represents a small fraction of the total number of stars within the volume element, $N=n\,V$. 
From Equation~(\ref{eq:Nb}), it is straightforward to show that the fraction of bound stars within the volume $V$ scales as $N_b/N\sim (v_e/\sigma)^3\,e^{-V_\bullet^2/(2\sigma^2)}$ at $r\gtrsim c_\bullet$, hence field particles can be treated independently from each other when the escape velocity is low, $v_e\lesssim \sigma$, or the substructure velocity is high, $V_\bullet\gg \sigma$.

The orbits of field stars captured by the potential $\Phi_\bullet$ exhibit chaotic fluctuations of energy and angular momentum, as shown in \S\ref{sec:ptt}. Statistically, one can count how many stars are bound to the substructure potential (i.e. $E<0$) at any given time and compare that number against the value estimated in Equation~(\ref{eq:Nb}).
 H\'enon \& Petit (1986) found that captured particles moving on chaotic orbits only remain trapped within the potential $\Phi_\bullet$ over a finite amount of time before being lost to galactic tides. As shown in Paper I, this leads to a population of bound field stars that reaches a {\it steady state} as the capture rate (defined as the net number of field particles with $E$ flipping from positive to negative values) equals the loss rate (i.e the net number of field particles with $E$ flipping from negative to positive values).
 Paper I finds that steady state is typically reached over Smoluchowski's (1916) `fluctuation mean life'
\begin{align}\label{eq:Ta}
  T(r)=\sqrt\frac{2\pi}{3}\frac{r}{\langle v^2\rangle^{1/2}},
\end{align}
which roughly corresponds to the time that a particle moving on a straight-line trajectory takes to cross the volume element, i.e. the so-called `crossing' time. Typically, this time-scale is much shorter than the orbital time of the substructure around the host galaxy. Paper I shows that in steady-state the average number of field stars with $E<0$ can be estimated from Equation~(\ref{eq:Nb}) as $N_{\rm ss}=\alpha\,N_b$, where $\alpha$ is the so-called abundance parameter, which is set by the `dynamical survival time' \footnote{Defined as the 
 time over which a trapped particle continuously has $E<0$ before being lost to galactic tides. } that trapped objects remain bound to the potential $\Phi_\bullet$. Numerical experiments shown in Paper I show that $\alpha$ is close to unity. The estimates shown below set the abundance parameter $\alpha=1$.

\subsubsection{Strong perturbations}\label{sec:strong}
The statistical theory presented above assumes that the substructure potential $\Phi_\bullet$ induces weak perturbations on the trajectories of field stars as they move across the volume $V=4\pi r^3/3$. One can easily show that this approximation breaks down close to the substructure, where the potential well $|\Phi_\bullet|$ may become deeper than the specific kinetic energy of field particles, $K=3\sigma^2/2$.

To derive the distribution of captured stars at small distances from the substructure we use two empirical results found in Paper~I (see also \S\ref{sec:experiment}): (i) the phase-space density of captured stars becomes approximately constant in the vicinity of the substructure, such that $f_\star({\bf r},{\bf v})\simeq f_0$, and (ii) the population of captured stars reaches a steady state on time scales comparable to Smoluchowski's (1916) `fluctuation mean life' defined by Equation~(\ref{eq:Ta}). 

 The density profile of an equilibrium ensemble of stars moving in the potential $\Phi_\bullet$ can be derived from the local distribution function without a priori knowledge of their orbital trajectories. That means, we do not need to specify whether orbits are regular ({\it permanently} bound) or chaotic ({\it temporarily} bound), as long as the distribution of these objects is homogeneous in phase-space. Under this condition,
\begin{align}\label{eq:rhoeq}
n_\star[\Phi_\bullet(r)]&=\int_{E<0} \d^3 v f_\star({\bf r},{\bf v}) \\ \nonumber
&= 4 \pi f_0\,\int_0^{|\Phi_\bullet|} \d E [2(E-\Phi_\bullet)]^{1/2}\\ \nonumber
&= \frac{8\sqrt{2}\pi}{3} f_0|\Phi_\bullet|^{3/2}.
\end{align}
To find the normalization of the local phase-space density, $f_0$, we match the number density~(\ref{eq:rhoeq}) against the steady-state profile derived from Equation~(\ref{eq:Nb}) at low escape velocities ($v_e\lesssim \sigma)$, such that $n_\star=\d\, N_{\rm ss}/\d^3 r=\alpha\,\d\,N_b/\d^3r$. This yields 
\begin{align}\label{eq:f0}
f_0=\alpha \frac{n}{(2\pi \sigma^2)^{3/2}}\,e^{-V_\bullet^2/(2\sigma^2)}.
\end{align}
which corresponds to a local Maxwellian distribution function multiplied by abundance parameter $\alpha$. The density enhancement induced by the population of energetically-bound stars within the volume $V$ can be found by inserting~(\ref{eq:f0}) into~(\ref{eq:rhoeq}), which yields
\begin{align}\label{eq:delta}
  \delta_\star(r)&\equiv \frac{n_\star(r)}{n}\\ \nonumber
 &=\alpha\frac{4}{3\sqrt{\pi}}e^{-V_\bullet^2/(2\sigma^2)} \frac{1}{\sigma^3}|\Phi_\bullet|^{3/2} \\ \nonumber
  &=\alpha\frac{4}{3\sqrt{\pi}}\frac{(G M_\bullet)^{3/2}}{\sigma^3} e^{-V_\bullet^2/(2\sigma^2)}\frac{1}{(r+c_\bullet)^{3/2}}.
\end{align}
Notice that the density profile converges to a centrally-divergent `density spike' $\delta_\star \sim r^{-3/2}$ (Gondolo \& Silk 1999) in the point-mass limit $c_\bullet\to 0$. For extended halos ($c_\bullet >0$) the profile becomes flat for $r\lesssim c_\bullet$. For generic potentials, we find that the density enhancement~(\ref{eq:delta}) scales as $\delta_\star\sim|\Phi_\bullet|^{3/2}$, which allows a straightforward analysis of a wide range of substructure models. 

The isotropic 1D velocity dispersion can be derived from the local distribution function as
\begin{align}\label{eq:vdisp}
\sigma^2_\star[\Phi_\bullet(r)]&=\frac{1}{3 \,n_\star(r)}\int_{E<0} \d^3 v \, v^2 f_\star({\bf r},{\bf v}) \\ \nonumber
&= \frac{4\pi \,f_0}{3\,n_\star(r)}\,\int_0^{|\Phi_\bullet|} \d E [2(E-\Phi_\bullet)]^{3/2}\\ \nonumber
&= \frac{2}{5}|\Phi_\bullet|,
\end{align}
which is a constant fraction of the escape velocity at all radii, $\sigma_\star/v_e=(1/5)^{1/2}\approx 0.45$. It is important to emphasize that Equation~(\ref{eq:vdisp}) is independent of the speed of the dark object across the host galaxy ($V_\bullet$) as well as of the normalization of the distribution function ($f_0$).

By construction, combining Equations~(\ref{eq:delta})  and~(\ref{eq:vdisp}) returns a mean phase-space density of bound stars that remains constant across the volume element
\begin{align}\label{eq:Qstar}
Q_\star\equiv \frac{n_\star(r)}{\sigma^3_\star(r)} =\alpha\frac{5}{3}\sqrt{\frac{10}{\pi}} e^{-V^2_\bullet/(2\sigma^2)}\,Q.
\end{align}
Equation~(\ref{eq:Qstar}) highlights two interesting results. First, $Q_\star$ solely depends on the potential $\Phi_\bullet$ through the value of the abundance parameter, $\alpha$, and second the phase-space density of captured stars drops exponentially for a potential $\Phi_\bullet$ that is not at rest with the background ($V_\bullet>0$).In \S\ref{sec:nbody}, we carry a number of numerical experiments that inspect the accuracy of the theoretical profiles derived above.

Of particular relevance for this paper is the distance at which the density of bounds stars is equal to that of the field. For reasons that become clear below, it is convenient to examine the point-mass limit first. Solving $\delta(r_\epsilon)=1$ in Equation~(\ref{eq:delta}) with $c_\bullet\to 0$ and $\alpha=1$ yields
\begin{align}\label{eq:r_eps}
  r_\epsilon=\bigg(\frac{16}{9\pi}\bigg)^{1/3}\,e^{-V_\bullet^2/(3\sigma^2)}\frac{GM_\bullet}{\sigma^2},
\end{align}
which is dubbed the `thermal' critical radius in Paper I owing to the Maxwellian velocity dependence $\exp[-V_\bullet^2/(3\sigma^2)]$ multiplying the critical radius $r_0=2 GM_\bullet/\sigma^2$ in~(\ref{eq:r_eps}). Thus, the density of field stars bound to an object with mass $M_\bullet$ will exceed that of the galactic background on scales below the thermal critical radius, i.e. $\delta_\star>1$ at $r<r_\epsilon$.

For objects with an extended mass distribution, $\delta_\star(r_\epsilon^{\rm ext})=1$ happens at $r_\epsilon^{\rm ext}=r_\epsilon-c_\bullet= \kappa\,r_\epsilon$, where $\kappa\equiv 1-c_\bullet/r_\epsilon$ is the {\it compactness} of the substructure. It is straightforward to show that the condition $\delta_\star>1$ requires $\kappa>0$ (or $c_\bullet < r_\epsilon$), which means that only substructures that are sufficiently {\it compact} can contain an over-dense region of captured field stars.
We inspect this issue in next Section with the aid of $N$-body tools.

In what follows, we assume that a dark substructure becomes `visible' --i.e. it can be detected as a localized stellar over-density-- if the volume $V_\epsilon=4\pi r_\epsilon^3/3>0$ contains at least one bound star, i.e. $N_b(r_\epsilon)> 1$. This condition can only be satisfied by dark objects above a certain mass threshold, $M_\bullet>M_{\rm min}$. Using~(\ref{eq:delta}) with $c_\bullet=0$ and~(\ref{eq:r_eps}) and setting $\alpha=1$ we find
\begin{align}\label{eq:Mmin}
  M_{\rm min}=\frac{3\pi^{1/3}}{4} \,e^{V_\bullet^2/(3\sigma^2)} \frac{D \,\sigma^2}{G},
\end{align}
here $D=(2\pi \,n)^{-1/3}$ is a measure of the average separation of stars in the volume element $V$ (Pe\~narrubia 2018). Notice again that the formation of stellar over-densities is strongly suppressed in dark objects moving at high speed ($V_\bullet \gg \sigma$) with respect to the galactic background.

\begin{figure}
\begin{center}
\includegraphics[width=84mm]{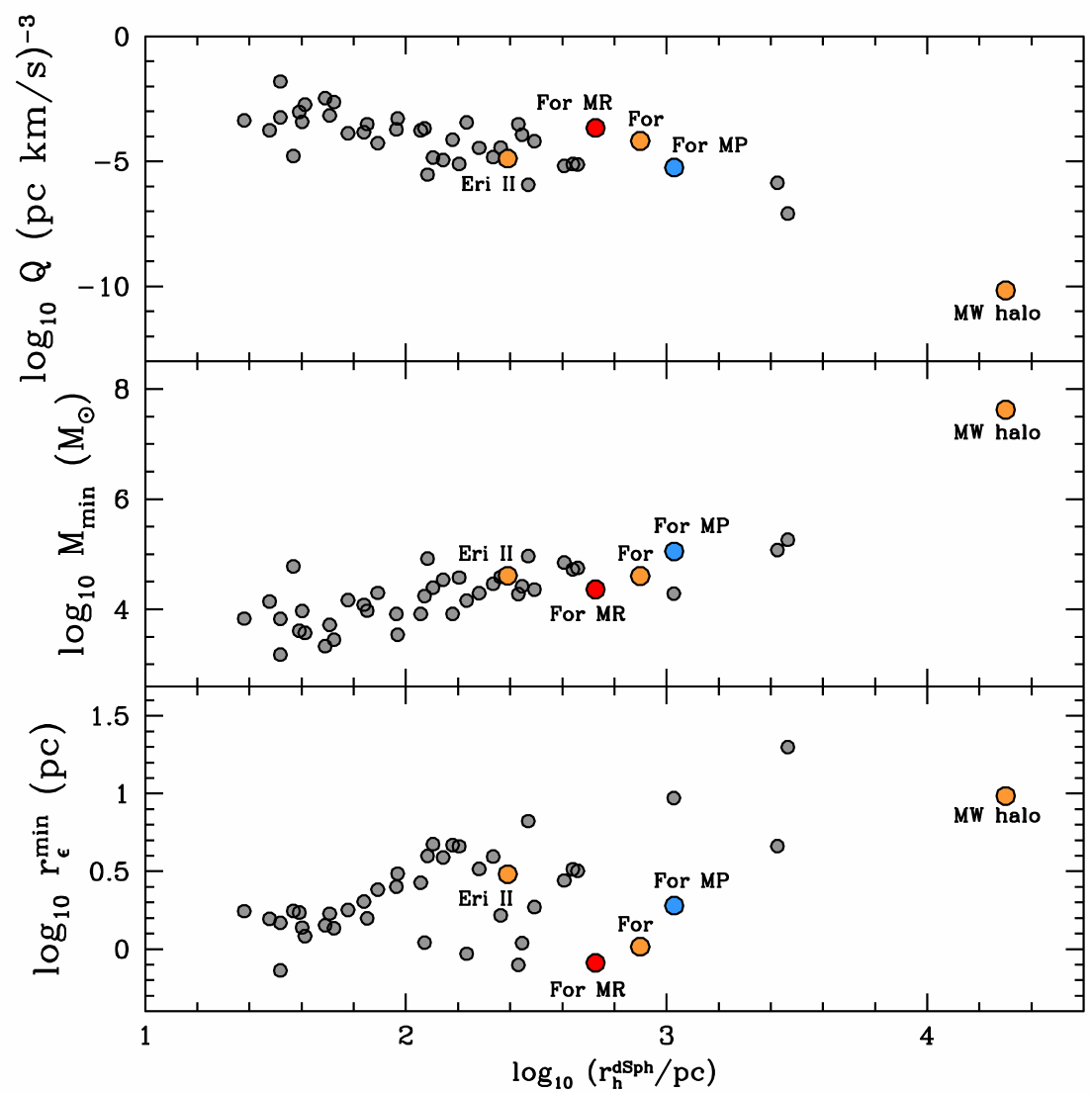}
\end{center}
\caption{{\it Upper panel}: Stellar phase-space density $Q=n/\sigma^3=1/(2\pi D^3\sigma^3)$ of MW dSphs as a function of their half-light radii. References for the data shown are given in footnote \ref{foot:dSphref}. Notice that the phase-space density of the MW stellar halo at $R=20\kpc$ is $\gtrsim 6$ orders of magnitude lower than in dSphs. This difference grows at larger Galactocentric radii.
{\it Middle panel}: Minimum substructure mass ($M_{\rm min}$) derived from Equation~(\ref{eq:Mmin}). Substructures with $M_\bullet>M_{\rm min}$ are expected to contain stellar over-densities ($\delta_\star>1$) with at least one bound field star ($N_b>1$), see text. The metal-rich (MR) and metal-poor (MP) components of the Fornax dSph are highlighted with red and blue filled dots, respectively. Notice that the value of $M_{\rm min}$ in the MW halo is relatively large on account of its high velocity dispersion. {\it Lower panel}: Minimum thermal critical radius $r_\epsilon^{\rm min}$ of substructures with a mass $M_\bullet=M_{\rm min}$. Stellar over-densities made of capture stars have a physical size comparable to the radii of globular clusters.}
\label{fig:mmin}
\end{figure}

\subsection{Application: Milky Way dwarf spheroidals}\label{sec:MW}
The above analytical expressions can be used to inspect the efficiency of gravitational captures in different galactic environments. 
According to our statistical theory, capture of field stars is most likely in regions where the stellar phase-space density $Q=n/\sigma^3=(2\pi D^3 \,\sigma^3)^{-1}$ is high, that is the intra-stellar separation $D$ is small and the velocity dispersion is low (i.e. `cold'). Upper panel of Fig.~\ref{fig:mmin} shows that the galaxies with the highest phase-space densities are the ultra-faint dwarf spheroidals, which also are the smallest, coldest and most metal-poor galaxies in the known Universe (e.g. Simon 2019). References for the data shown are given in footnote \ref{foot:dSphref}\footnotetext{\label{foot:dSphref}Half-light radii and velocity dispersion for the dSphs shown in Fig.~\ref{fig:mmin} are from McConnachie (2012) (version January 2021), with updates for Antlia~2, Crater~2 (Ji et al. 2021) and Tucana (Taibi et al. 2020). }.

In what follows, we assume that the stellar component of satellite galaxies follow a Plummer profile, $n^{\rm dSph}(r)=n^{\rm dSph}_0[1+(r/a)^2]^{-5/2}$, with a central density $n^{\rm dSph}_0=N^{\rm dSph}_\star/(4\pi a^3/3)$, and a (3d) half-light radius $r^{\rm dSph}_h=1.305\,a$. The average inter-stellar separation is estimated as $D=(2\pi n_0^{\rm dSph})^{-1/3}$. The number of stars in a dSph galaxy is calculated from the total stellar mass as $N^{\rm dSph}_\star=M^{\rm dSph}_\star/\langle m_\star\rangle $, where, $\langle m_\star\rangle$ is the average mass of a single star. Given that dSphs have approximately flat velocity dispersion profiles, we set $\sigma=\sigma^{\rm dSph}$ (Walker et al., 2007). 

Dwarf galaxies with multiple chemo-dynamical components are particularly interesting in this context. In these systems, the probability to capture field stars strongly depends on stellar metallicity. For example, the Fornax dSph contains two prominent chemo-dynamical populations (Battaglia et al. 2006; Walker \& Pe\~narrubia 2011; Amorisco \& Evans 2012) with metallicities $\mathrm{[Fe/ H]}= -1.8$ and $-0.65$ and ages $>10\gyr$, and $\sim 0.2$--$2\gyr$, respectively (Rusakov et al. 2021). The metal-rich (MR) component contains more stars $N_\star^{\rm MR}\sim 3\times 10^7$, is more centrally concentrated, $r_h^{\rm MR}\approx 530\pc$ and has a colder velocity dispersion, $\sigma^{\rm MR}\approx 10\kms$, than the metal-poor (MP) population, which has $N_\star^{\rm MP}\sim 2\times 10^7$, $r_h^{\rm MP}\approx 1070\pc$ and $\sigma^{\rm MP}\approx 14.4\kms$ (Walker \& Pe\~narrubia 2011). Thus, MR stars have much higher phase-space densities than MP stars, 
\begin{align}\label{eq:Qrat}
\frac{Q^{\rm MR}}{Q^{\rm MP}}= \frac{n^{\rm MR}}{n^{\rm MP}}\bigg(\frac{\sigma^{\rm MP}}{\sigma^{\rm MR}}\bigg)^3=\frac{N_\star^{\rm MR}}{N_\star^{\rm MP}}\bigg(\frac{r_h^{\rm MP}}{r_h^{\rm MR}}\frac{\sigma^{\rm MP}}{\sigma^{\rm MR}}\bigg)^3\sim 33.
\end{align}

Middle panel of Fig.~\ref{fig:mmin} shows estimates of $M_{\rm min}$ derived from Equation~(\ref{eq:Mmin}) for the MW satellites in our sample. To simplify the analysis, substructures are placed at rest within the host galaxies ($V_\bullet=0$). For reference, we also plot the value of $M_{\rm min}$ associated with the stellar halo of the Milky Way at a Galactocentric distance of $R=20\kpc$ using the stellar density and velocity dispersion measured by Deason et al. (2011).
This panel reveals a number of interesting points. Notice first that the mass threshold ($M_{\rm min}$) is lowest in the smallest dSphs, which simply reflects the tight correlation between half-light radius and phase-space density shown in the upper panel. If the goal is to detect `dark' substructures that contain a stellar population of captured field stars, then ultra-faint dSphs with half-light radii $r_h^{\rm dSph}\sim 30\pc$ are the best targets, as they may be sensitive to the presence of dark substructures with masses above $M_\bullet\gtrsim 10^4\msol$. In contrast, the mass threshold increases by an order of magnitude, $M_\bullet\gtrsim 10^5\msol$, in the `classical' dwarf galaxies with $r_h^{\rm dSph}\gtrsim 300\pc$. In the case of the Fornax dSph, 
the MR stellar component a mass threshold $M_{\rm min}$ that is approximately an order of magnitude lower than for the MP population.

Stars trapped in dark substructures can lead to the formation of localized stellar overdensities ($\delta_\star>1$) in the host galaxy. Bottom panel of Fig.~\ref{fig:mmin} shows that the size of the overdense regions lies in the range $r_\epsilon\gtrsim 1$--$10\pc$, thus being comparable to the size of stellar clusters. This leads to the intriguing possibility that some of the `anomalous' stellar systems detected in MW dSphs may actually correspond to stars temporarily trapped in dark substructures. We discuss this scenario in \S\ref{sec:data}.

Overdensities of trapped stars are expected to have high mass-to-light ratios. This can be shown by measuring the substructure mass within the over-density volume as $M=M_\bullet(<r_\epsilon)=M_\bullet r_\epsilon^2/(r_\epsilon+c_\bullet)^2$, and comparing it to the bound stellar mass $L=N_b(r_\epsilon)\, \langle m_\star\rangle $. For the point-mass case $c_\bullet=0$, applying~(\ref{eq:Nb}) and~(\ref{eq:r_eps}) returns a dimension-less mass-to-light ratio 
\begin{align}\label{eq:ML}
  \frac{M}{L}\equiv \frac{M(<r_\epsilon)}{N_b(r_\epsilon)\,\langle m_\star\rangle }=\frac{M_{\rm min}^3}{M^2_\bullet\,\langle m_\star\rangle },
\end{align}
where $M_{\rm min}$ is given by equation~(\ref{eq:Mmin}). According to the estimates plotted in Fig.~\ref{fig:mmin}, field stars bound to susbtructures with a mass $M_\bullet\sim M_{\rm min}$ will appear as extremely DM-dominated objects, $M/L= M_{\rm min}/\langle m_\star\rangle \sim 10^4$--$10^5$ for an average stellar mass $\langle m_\star\rangle=1\msol$. In substructures with larger masses, $M_\bullet> M_{\rm min}$, the mass-to-light ratio of captured stars drops as $M/L\sim M_\bullet^{-2}$. Note that Equation~(\ref{eq:ML}) should not be applied to substructures where $M/L\lesssim 1$, as the theory outlined in \S\ref{sec:number} ignores the contribution of trapped stars to the underlying potential.

In the MW halo, capture of field stars is inefficient on account of their high velocity dispersion. For illustration, here we adopt $\sigma^{\rm MW}\sim 124\kms$ and $n^{\rm MW}=10^{-4}\pc^{-3}$ at $R=20\kpc$ from the MW centre (e.g. Deason et al. 2011), noting in passing that capture of field stars in this particular stellar halo model becomes systematically less efficient at larger distances. 
Note that the velocity dispersion of the MW stellar halo is approximately one order of magnitude higher than in the classical dSphs.
As a result, the minimum substructure mass needed to capture stars from the field increases up to $M_{\rm min}\gtrsim 10^8\msol$, which is $\gtrsim 3$ orders of magnitude larger than in the MW dwarf spheroidals. 
Crucially, this mass threshold is lower than the virial mass of the classical dSphs, which appear to be embedded in DM haloes with virial masses $10^9$--$10^{10}\msol$ (e.g. Pe\~narrubia et al. 2008; Errani et al. 2018), suggesting that dSphs may contain a population of captured MW halo stars. However, field haloes in this mass range have scale radii that are a factor $\gtrsim 100$ larger than the thermal critical radius plotted in the bottom panel of Fig.~\ref{fig:mmin}, $c_\bullet\gtrsim 10^3\pc$ (e.g. Joyce \& Diemand 2019), which implies a negative compactness parameter, $\kappa<0$. Therefore, field stars trapped in dSphs will appear as a diffuse envelope of co-moving stars with subdominant densities with respect to the local MW background ($\delta_\star\ll 1$). The detection of this stellar population will be challenging, and may require simultaneous modelling of the kinematic and chemical composition of a large sample of stars at the locations of dSphs. This problem will be explored in a separate contribution.

\begin{figure*}
\begin{center}
\includegraphics[width=170mm]{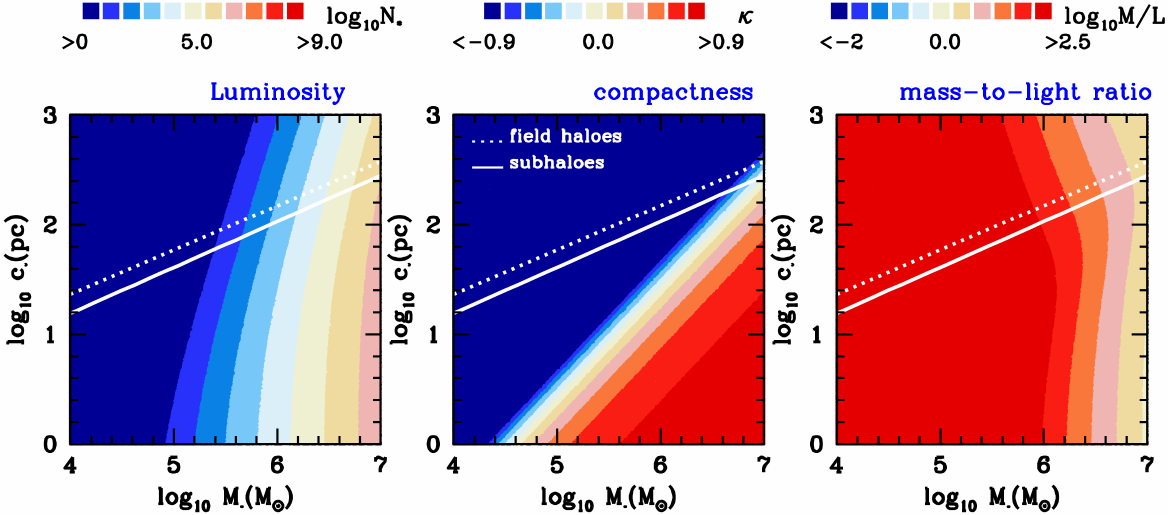}
\end{center}
\caption{{\it Left panel:} mass $M_\bullet$ and scale radius $c_\bullet$ of DM sub-subhaloes colour-coded according to the number of captured dwarf galaxy stars, $N_\star=N_b(r_\epsilon)$, Equation~(\ref{eq:Nb}), where $r_{\epsilon}$ is the thermal critical radius, Equation~(\ref{eq:r_eps}). In this model, the host galaxy has properties similar to the Fornax dSph: it follows a Plummer profile with a total luminosity of $N^{\rm dSph}=5\times 10^7$, a half-light radius $r_h^{\rm dSph}=0.89\kpc$ and it has an average velocity dispersion $\sigma^{\rm dSph}=12.2\kms$. White solid and dotted lines respectively show the virial mass-size relation of haloes and subhaloes taken from the Aquarius simulations (Springel et al. 2008, fig. 26) extrapolated down to the mass scales of interest (see text). {\it Middle panel} Compactness $\kappa=1-c_\bullet/r_{\epsilon}$ of the left-panel models. Only models with $\kappa>0$ lead to over-densities of captured stars with $\delta_\star>1$. {\it Right panel}: Mass-to-light rations of the left and middle-panel models derived from Equation~(\ref{eq:ML}). Stellar systems trapped in subhaloes with masses $M_\bullet\lesssim 10^7\msol$ will exhibit dark-matter dominated kinematics. }
\label{fig:kml}
\end{figure*}
\subsection{Sub-subhalo populations in a CDM framework}\label{sec:sub}
According to CDM models of structure formation, all galaxies are expected to contain a large population of DM subhaloes with a power-law mass function $\d N/\d M_\bullet\sim M_\bullet^{-\alpha}$ and a slope $\alpha\approx 1.9$ (e.g. Giocoli et al. 2008; Springel et al. 2008). In the classical fluid limit, the number of subhaloes 
diverges at low masses. In contrast, if DM is made of cold massive particles a truncation at a 
 minimum subhalo mass $M_{\rm min}$ is expected to arise on scales comparable to the particle free-streaming length. For WIMPS with masses above $\sim 1 \mathrm{GeV/c}^2$ the truncation lies below planet-mass scales, $M_{\rm min}\lesssim 10^{-6}\, \rm M_\odot$  (e.g. Schmid et al. 1999; Hofmann et al. 2001). 

 Within a CDM context, dark subhaloes hosting a population of captured field stars are expected to follow a luminosity function dictated by the underlying halo mass function. To illustrate this, let us define $N_\star=N_b(r_{\rm out})$ within a volume size larger than the substructure scale radius, $r_{\rm out}\gg c_\bullet$. From Equation~(\ref{eq:Nb})  
\begin{align}\label{eq:Nb_pm}
N_\star=\frac{32\sqrt{\pi}}{9}(G M_\bullet)^{3/2}\,e^{-V_\bullet^2/(2\sigma^2)}\frac{n}{\sigma^3} \, r_{\rm out}^{3/2},~~~~{\rm for}~~~ r_{\rm out}\gg c_\bullet.
\end{align} 
To combine Equation~(\ref{eq:Nb_pm}) with the subhalo mass function we apply the chain rule $(\d N/\d M_\bullet)\d M_\bullet \sim N_\star^{-(2\alpha+1)/3}\d N_\star$, which leads to a luminosity function
\begin{align}\label{eq:lumfunc}
\frac{\d N}{\d M_\star}=A_0 \bigg(\frac{M_\star}{\langle m_\star\rangle}\bigg)^{-(2\alpha+1)/3},
\end{align}
where $M_\star=N_\star \langle m_\star\rangle$ is the mass of a stellar clump, and $A_0$ is an arbitrary normalization. Thus, the luminosity function of stellar substructures also follows a power-law $\d N/\d M_\star\sim M_\star^{-\beta}$, albeit with a shallower index than the underlying subhalo mass function, $\beta=(2\alpha+1)/3=1.6$ for $\alpha=1.9$.

Whether or not the these objects can be detected as localized stellar over-densities mainly depends on the {\it compactness} of DM haloes that capture them (see \S\ref{sec:model}).
To illustrate this point, Fig.~\ref{fig:kml} shows the mean properties of field stars trapped in dark substructures covering a wide range of masses and scale radii. For this plot, we adopt the same dwarf galaxy model as outlined in \S\ref{sec:nbody}. Namely, field stars follow a Plummer profile with a total luminosity of $N^{\rm dSph}=5\times 10^7$, a half-light radius of $r_h^{\rm dSph}=0.89\kpc$ and a mean velocity dispersion $\sigma^{\rm dSph}=12.2\kms$, similar to the overall properties of the Fornax dSph (see \S\ref{sec:MW}). For simplicity, we assume that dark substructures are at rest with the field ($V_\bullet=0$).

Left panel of Fig.~\ref{fig:kml} shows models colour-coded according to the number of bound stars enclosed within their thermal critical radius, $N_b(r_{\epsilon})$ calculated from Equations~(\ref{eq:Nb}) and~(\ref{eq:r_eps}). As shown in Fig.~\ref{fig:mmin}, point-mass objects with $M_\bullet\gtrsim 10^5\msol$ contain $N_\star\gtrsim 1$ stars trapped within their thermal critical radius. In extended substructures, the mass threshold for capture increases mildly.

Only substructures with positive {\it compactness} ($\kappa\equiv 1-c_\bullet/r_\epsilon$) can become visible as localized stellar over-densities ($\delta_\star>1$). Middle panel shows that the condition $\kappa>0$ (or $c_\bullet<r_\epsilon$) translates into a linear relation $c_\bullet\lesssim 0.83\,GM_\bullet/\sigma^2$, which is steeper than the shallow mass-scale radius relation of Aquarius subhaloes, $c_\bullet\sim M_\bullet^{0.46}$. As a result, Aquarius subhaloes with masses $M_\bullet\lesssim 5\times 10^6 \msol$ would be `fluffy' ($\kappa<0$). In these objects, trapped stars would have sub-dominant densities with respect to the field ($\delta_\star<1$), complicating their detection in photometric surveys. However, in spectroscopic surveys these objects may appear as localized regions with distinct kinematics, i.e. a low velocity dispersion $\sigma_\star<\sigma$, and/or a significant velocity offset $\Delta v$.

It is important to stress that the mass-size relation of field haloes and subhaloes contain a significant amount of scatter that is not plotted in Fig.~\ref{fig:kml} (e.g. the concentration of field haloes has a standard deviation of $\sim 0.15\,$dex at a fixed mass, Ludlow et al. 2016).
Crucially, a low-probability, high-density tail in the mass-size distribution would lead to a population of low-luminosity over-densities with the size of stellar clusters. This is of particular relevance as the existence of such a high-density, low-probability tail of substructures is a strong prediction from the cold dark matter paradigm (Pe\~narrubia et al. 2010; Errani \& Pe\~narrubia 2020; Errani \& Navarro 2021).
Unfortunately, given the poor theoretical understanding of the mass function and the density profile of sub-subhaloes in dSphs, at present it is not possible to make predictions on the {\it number} of stellar over-densities of captured field stars. We will discuss this issue in \S\ref{sec:data}.

Right panel of Fig.~\ref{fig:kml} shows the mass-to-light ratio~(\ref{eq:ML}) for the models shown in previous panels. As expected, we find that field stars trapped in low-mass ($M_\bullet\lesssim 10^7 \msol$) substructures exhibit DM-dominated kinematics ($M/L\gg 1$). As a note of caution, recall that the analytical equations derived in \S\ref{sec:number} neglect the self-gravity of captured stars, which is not a valid assumption in systems with mass-to-light ratios $M/L\lesssim 1$.

In summary, Fig.~\ref{fig:kml} suggests that field stars captured by dark sub-subhaloes may resemble stellar clusters with anomalous properties: (i) they would have a large size for their luminosity, (ii) contain stellar populations with ages and metallicities indistinguishable from the host galaxy, and (iii) exhibit DM-dominated mass-to-light ratios.


\section{Statistical Experiments }\label{sec:experiment}
This Section carries a suite of numerical experiments that help us to test the accuracy of the theoretical equations derived in \S\ref{sec:number}, and shed light on the dynamics of stars trapped by a substructure orbiting in a Fornax-like dSph galaxy. The Section goes as follows

\S\ref{sec:nbody} provides a brief overview of the properties of the galactic field particles and the numerical set-up used to solve the equations of motion. We refer interested readers to Appendix A of Paper~I for a detailed description of the integration tools.

\S\ref{sec:ptt} injects a static substructure potential $\Phi_\bullet$ in a pre-existing field of particles in dynamical equilibrium within $\Phi^{\rm dSph}$. This set-up is similar to the analytical conditions adopted in \S\ref{sec:number}, and we can therefore anticipate a close match between theoretical predictions and the numerical results. Interestingly, we will see that the dynamics of field particles trapped in $\Phi_\bullet$ can be broadly separated in two families: (i) {\it permanent} captures, which become bound immediately after the substructure is placed in the field at $t=0$, and remain bound for indefinitely long times, and (ii) {\it temporary} captures, which as the name indicates only remain bound for a finite amount of time.

\S\ref{sec:growth} explores a physically-motivated case
where a substructure mass {\it grows} while moving along a circular orbit, thus sourcing a time-dependent potential $\Phi_\bullet(t)$. We will see that these models also capture particles trapped on {\it permanent} orbits around $\Phi_\bullet$, and that one time-scales $t\gtrsim |\dot \Phi_\bullet/\Phi_\bullet|^{-1}$ the phase-space distribution of particles with $E<0$ become indistinguishable from the static case explored in \S\ref{sec:ptt}.

In \S\ref{sec:accretion} we analyze capture of field stars by a substructure accreted onto a dwarf galaxy on an eccentric orbit. The potential $\Phi_\bullet$ is initially placed at orbital apocentre in a region populated by no field particles (therefore it contains no permanent captures by construction), with a small orbital pericentre that reaches the inner-most regions of the dSph.

\S\ref{sec:fluffy} considers substructure models with a fixed mass and different scale radii. The numerical results stress that only substructures that are sufficiently compact generate overdensities of captured field stars.

\subsection{Initial conditions \& set-up}\label{sec:nbody}
We generate realizations of $N=5\times 10^6$ stellar tracer particles in dynamical equilibrium within a Dehnen (1993) potential $\Phi^{\rm dSph}$ with a total mass of $M^{\rm dSph}=3\times 10^9\msol$ and a scale radius $c^{\rm dSph}=2\kpc$. We run experiments with cuspy ($\gamma=1$) and cored ($\gamma=0$) profiles.
Given that the stellar luminosity of the Fornax dSph is $\sim 5\times 10^7\lsol$ (e.g. Rusakov et al. 2021), the particle luminosity in our models is $10\lsol$. 

Field particles in this potential follow an $\alpha$-$\beta$-$\gamma$ profile (Zhao 1996)
\begin{align}\label{eq:n_abg}
n(R)=\frac{n_0}{(R/R_0)^{\gamma_f}[1+(R/R_0)^{\alpha_f}]^{(\beta_f-\gamma_f)/\alpha_f}},
\end{align}
with $n_0$ chosen such that $4\pi\int_0^\infty\d r\, r^2\, n(r)=N$. We consider two tracer models: (i) a
spherical Plummer-like (1911) profile\footnote{We notice in order to generate a positive distribution function via an Eddington inversion, the inner slope $\gamma_f$ can be small but not exactly zero.} with a scale radius $R_0=690\pc$ and slopes $(\alpha_f,\beta_f,\gamma_f)=(2,5,0.1)$, and (ii) a truncated model with  $(\alpha_f,\beta_f,\gamma_f)=(2,30,0.2)$ and a scale radius $R_0=460\pc$. 
To guarantee dynamical equilibrium, orbital velocities are assigned using Eddington (1916) inversion (see Errani \& Pe\~narrubia 2020). For the Plummer profile, this returns a luminosity-averaged 1D velocity dispersion $\langle v^2\rangle^{1/2}/3=12.2\kms$. This is our reference model, which approximately matches the overall phase-space density, size and luminosity of the Fornax dSph plotted in Fig.~\ref{fig:mmin}. The luminosity-averaged velocity dispersion of the truncated model is $\langle v^2\rangle^{1/2}/3=11.7\kms$.

The motion of individual tracer particles are solutions to two sets of differential equations. The first set describes the orbit of the substructure in the dSph potential
 \begin{align}\label{eq:eqmotsun}
 \ddot {\bb R_\bullet}=-\nabla\Phi^{\rm dSph}({\bb R_\bullet}).
 \end{align}
 
 The second set describes the trajectories of tracer particles in the galactocentric frame
 \begin{align}\label{eq:eqmot}
   \ddot {\bb R}=-\nabla\Phi^{\rm dSph}(\bb R)-\nabla\Phi_\bullet(\bb R-\bb R_\bullet) + {\bb F}_{\rm coll},
 \end{align}
where $\bb R-\bb R_\bullet$ is the relative distance of the particle to the substructure, and ${\bb F}_{\rm coll}=\sum_{i=1}^{N_{\rm clump}} {\bb f}_i$ is the net force induced by$N_{\rm clump}$ clumps orbiting in the substructure potential $\Phi_\bullet$ (for example, these could be planets orbiting the Sun in Paper I, or sub-subhaloes in the DM halo of dSphs in the current work). For simplicity, here we assume that the substructure potential is `smooth' by setting $F_{\rm coll}=0$. We will analyze the effect of random encounters with clumps on the dynamics of captured particles in follow-up work.

In Sections~\ref{sec:ptt} and~\ref{sec:growth}, substructures are placed on circular orbits around a cored DM halo ($\gamma=0$) at a galactocentric radius $R_\bullet=0.5\kpc$ with a tangential velocity $V_\bullet=V_c(R_\bullet)=14.35\kms$. The local dynamical time is therefore $\Omega^{-1}=R_\bullet/V_\bullet=500\pc/14.35\kms\simeq 34\myr$.
Field stars follow a Plummer (1911) profile, have a local velocity dispersion of $\sigma(R_\bullet)=11.2\kms$, and are separated by an average distance $D(R_\bullet)=4.8\pc$.

For illustration, we choose a substructure mass $M_\bullet= 10^6\msol$. To estimate the thermal critical radius we adopt $\sigma=\langle v^2\rangle^{1/2}/3=12.2\kms$ in Equation~(\ref{eq:r_eps}), which returns $r_\epsilon=18.9\pc$. 
Note that this size is comparable to that of the extended stellar clusters discussed in the Introduction. 
We consider a `compact' substructure with $\kappa=+0.8$, which translates into a scale radius $c_\bullet=(1-\kappa)\,r_\epsilon=3.78\pc$ (notice that this value is smaller than the typical scale radius of field haloes with a vitial mass $M_{\rm vir}(z=0)\sim 10^6 M_\odot$ in the Aquarius simulation, see Fig.~\ref{fig:kml}). 
The time-scale needed to cross the over-density size by a random field particle is much shorter than the orbital time,
$T(r_\epsilon)=r_\epsilon/\sigma=18.9\pc/(12.2\kms)\simeq 1.5\myr$. As a result, in these models the population of captured particles reaches steady state quickly.

We choose a volume size around $\Phi_\bullet$ which matches the local tidal radius derived below from Equation~(\ref{eq:rt}), $r_{\rm out}=r_t\simeq 190\pc$. The volume $V=4\pi r^3_{\rm out}/3$ contains approximately $N=n\,V=42530$ field stars. From Equation~(\ref{eq:Nb}), we expect that approximately $N_b(r_{\rm out})=2000$ stars will have  negative energies.

\begin{figure}
\begin{center}
\includegraphics[width=84mm]{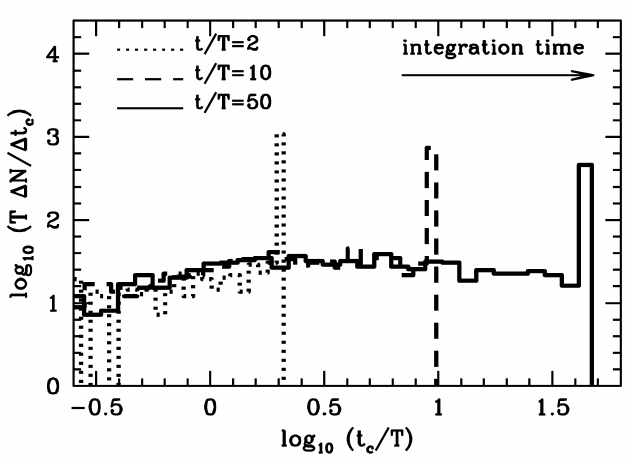}
\end{center}
\caption{ Distribution of capture times ($t_c=t-t_{\rm bound}$) defined as the time spanned since a particle becomes bound ($t_{\rm bound}$), i.e. the energy $E$ flips from positive to negative, until the present integration time ($t$). Time is measured in units Smoluchowski's time scale ($T$) for convenience. Here, we only include particles within a volume size equal to the local tidal radius, $r_{\rm out}=r_t$ (see text). Notice the presence of a substantial number of particles that become bound at $t\approx 0$, which result in a a strong peak in the distribution at $t_c\approx t$ shifting towards the right as the integration time increases.
}
\label{fig:time}
\end{figure}
\subsection{Static substructures potential on circular orbits}\label{sec:ptt}
In this Section we place a substructure potential $\Phi_\bullet$ at a galactocentric radius $R_\bullet$ with a circular velocity $V_\bullet=V_c(R_\bullet)$ in a sea of tracer particles with an extended profile (see \S\ref{sec:nbody}) in dynamical equilibrium. By chance, at the initial snapshot ($t=0$) a number of tracer particles are found within a volume element $V=4\pi r^3_{\rm out}/3$ centred at $\Phi_\bullet$ with a relative velocity below the escape velocity, $v<v_e$, which implies a negative energy, $E<0$. In what follows, we refer to these particles as `immediate captures' in order to distinguish them from `chaotic 3-body captures' (e.g. Petit \& Henon 1986), which take place during the dynamical integration of their orbits as a result of the interplay between the tracer particle and the substructure and host galaxy potentials.

Next, we follow the motions of the substructure and the field particles by solving Equations~(\ref{eq:eqmotsun}) and~(\ref{eq:eqmot}). At each time-step, we identify particles located within the volume $V$ whose energy flips from positive to negative, label them as `captured', and mark their ID's and the time when this happens ($t_{\rm bound})$. 

As the simulation proceeds, we notice that orbits with $E<0$ can be broadly separated into two families. The first group is made of `permanent' captures,
which correspond to particles that remain bound for arbitrarily long times and were already bound at $t=0$ (i.e. these are immediate captures that survive until the final snapshot of the simulation). Here, it is worth stressing that not all immediate captures are permanently bound, as some of them become tidally stripped by the galactic potential. We will come back to this point below. The second group corresponds to `temporary' captures. As the name indicates, these particles were unbound ($E>0$) at the initial snapshot $t=0$. At some point during the integration their energy sign flips from positive to negative, and back to positive, as these particles only remain bound for a finite amount of time. 

The presence of particles on permanently-bound orbits can be easily identified in the distribution of capture times plotted in Fig.~\ref{fig:time} at three different snapshots of the simulation. Independently of the time of the measurement, we find a very prominent spike centred at $t_c=t-t_{\rm bound}=t$. This spike is populated by particles that were already bound at $t=0$ ($t_{\rm bound}=0$). Away from the spike, we observe a broad distribution of capture times, which is populated by particles that were initially unbound and become captured at a later time of the simulation, such that $t_{\rm bound}>0$. Their distribution of capture times peaks at $t_c\sim 2\,T$, although in this experiment we also find a significant number of temporary captures that remain bound for as long as $t_c\sim 50\,T$. In our models, all particles captured with $t_{\rm bound}>0$ are temporary events that ultimately lead to the tracer particle being released back to the galactic potential.

\begin{figure}
\begin{center}
\includegraphics[width=84mm]{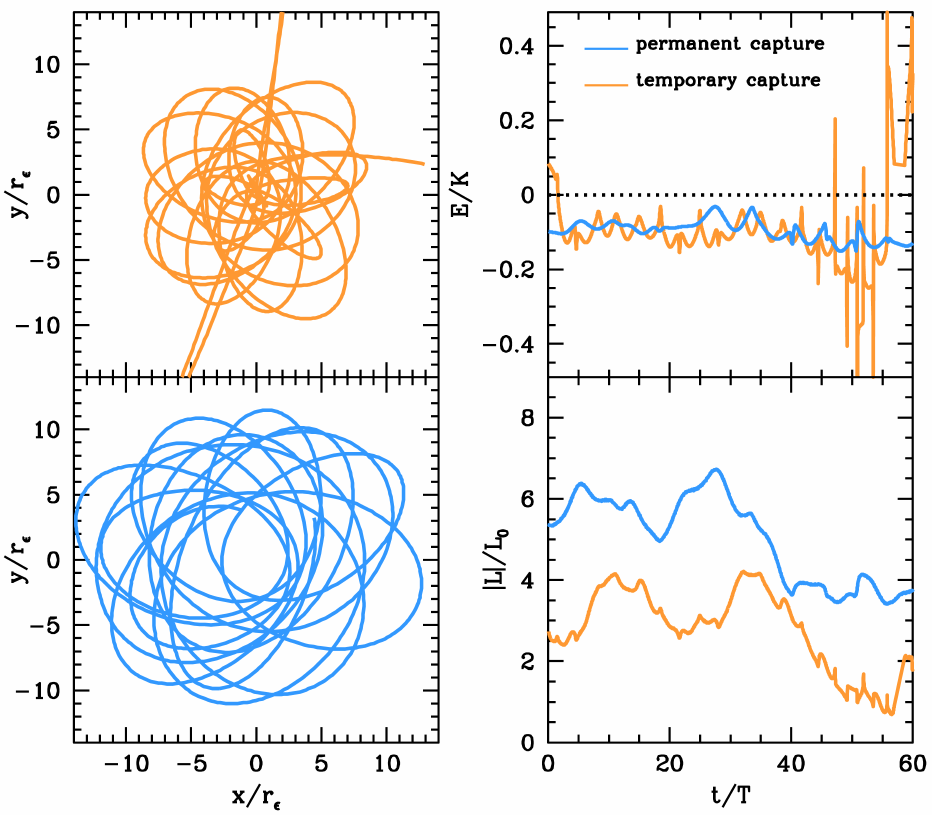}
\end{center}
\caption{Examples of permanent and temporary captures. {\it Left panels:} Motion of captured particles in a coordinate frame centred at the potential $\Phi_\bullet$ of a `compact' Hernquist substructure with $M_\bullet=10^6\msol$ and a compactness $\kappa=0.2$ (see \S\ref{sec:number} for details), which leads to a scale radius $c_\bullet=(1-\kappa)\,r_\epsilon=3.78\pc$. Notice how the temporary capture eventually leaves the volume of observation as it becomes gravitationally unbound.
{\it Upper and lower right panels}: Time-evolution of the specific energy $E=v^2/2+\Phi_\bullet$ and angular momentum $|\bb L|=|\bb r\times \bb v|$, respectively. Energy is measured in units of the mean kinetic energy of field stars, $K=3\sigma^2/2$, and angular momentum in units of $L_0=r_\epsilon\,\sigma$. Time is measured in units of Smoluchowski's (1916) fluctuation mean-life~(\ref{eq:Ta}). In contrast to temporary captures, particles on permanent orbits exhibit negative binding energies ($E<0$) during the entire time span of the integration.}  
\label{fig:orb}
\end{figure}

Since we know the particle ID's, we can follow the trajectories of permanent and temporary captures individually. For illustration, Fig.~\ref{fig:orb} shows the motion of two field particles captured by the potential $\Phi_\bullet$, with orange and blue lines showing temporary and permanent captures, respectively.
Temporary captures are transient events that inevitably result in the field particle escaping from the potential $\Phi_\bullet$ after a finite amount of time. In the example plotted in Fig.~\ref{fig:orb}, the specific energy of the field particle $E=v^2/2+\Phi_\bullet$ flips from positive to negative $t\simeq 2\,T$ and remains negative until $t\simeq 47\,T$, which yields a survival time $t_{\rm surv}\sim 45\,T$. After $t\gtrsim 47\,T$, the particle goes through a short period of rapid energy oscillations in which the orbit becomes bound and unbound repeatedly over short time-scales, $t_{\rm surv}\sim T$, until it finally becomes fully unbound and escapes from the volume element $V$ at $t\sim 60\,T$. 
In contrast, permanent captures have $E<0$ at all times and show milder variations of energy.
It is important to emphasize that none of these particles conserve energy or angular momentum ($\bb L=\bb r\times \bb v$),  which greatly complicates an analytical description of their motion. For example, their orbits are not typically restricted to a constant orbital plane --even when $\Phi_\bullet$ is spherical \& static-- and exhibit time-varying orbital parameters (such as peri, apo-centres and orbital eccentricity) that change in an apparently random fashion over short time scales.

\begin{figure}
\begin{center}
\includegraphics[width=84mm]{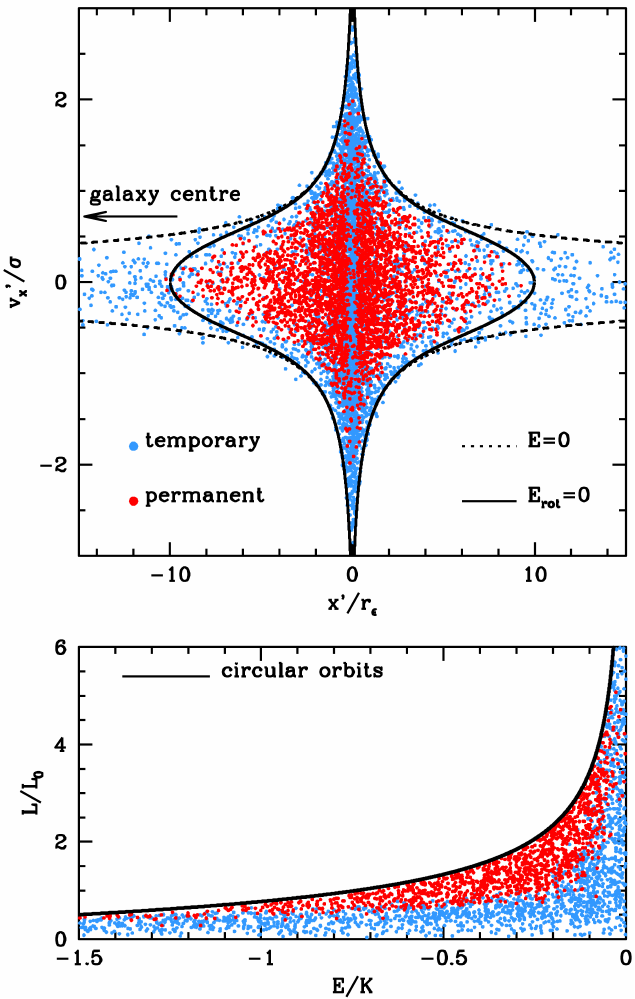}
\end{center}
\caption{{\it Upper panel:} Radial phase-space location of bound particles in a co-rotating frame, $x'$ -- $v_x'$ measured at the end of the simulation ($t=t_f$). Only particles within a narrow plane $|\Delta y'|<0.2\,r_\epsilon$ and $|\Delta z'|<0.2\,r_\epsilon$ are shown here. The dwarf galaxy centre is located at $x'/r_\epsilon\approx -25$. We identify `permanent' captures as those particles that remain bound for the whole span of the simulation $t_c=t_f$ (or $t_{\rm bound}=0$) and plot them with red dots, whereas temporary captures become bound at $t_{\rm bound}>0$ and are plotted in blue. Note that permanent captures have negative binding energies measured in a co-rotating frame, $E_{\rm rot}<0$, and are confined within the tidal radius $|x'|<r_t\simeq 10\,r_\epsilon$, with $r_t$ given by Equation~(\ref{eq:rt}). Notice also the lack of permanent captures at very small distances from the substructure, $r\ll r_\epsilon$. {\it Lower panel:} Energy and angular momentum of the particles plotted above given in the same units as in the right panels of Fig.~\ref{fig:orb}. Black-solid line shows the circular velocity of the Hernquist substructure. Notice that permanent and temporary particles appear segregated in the integral-of-motion space. In particular, temporary captures can be rarely found on circular orbits. Note also that this division blurs close to the 'fringe' region, $E\approx 0$.
}
\label{fig:rotate}
\end{figure}

As noted above, only a fraction of particles with $E<0$ at $t=0$ (the so-called immediate captures) remain bound for an indefinite amount of time. Fig.~\ref{fig:rotate} shows that only immediate captures that were initially located within the local tidal radius become permanently bound. This plot shows particles in a co-rotating coordinate system centred at the substructure, with the $x'$-axis pointing at the host galaxy centre, and the $z'$-axis aligned with the angular vector ${\bb \Omega}=(0,0,\Omega)$. In this frame, the effective potential can be written as (c.f. Eq.~12 of Renaud et al. 2011)
\begin{align}\label{eq:phi_eff}
\Phi_{\rm eff}=\Phi_\bullet -\frac{1}{2}(\lambda_1\,x'^2+\lambda_2\,y'^2+\lambda_3\,z'^2),
\end{align}
where $\lambda_i$ are the 3 eigen-values of the local tidal tensor. For a Dehnen (1993) sphere with $\gamma=0$ they can be expressed analytically using Eq.~(17) of Renaud et al. (2011) as 
\begin{align}\label{eq:lambda}
(\lambda_1,\lambda_2,\lambda_3)=\frac{GM^{\rm dSph}}{(R_\bullet+c^{\rm dSph})^3}\bigg(\frac{3 R_\bullet}{R_\bullet+c^{\rm dSph}},0,-1\bigg).
\end{align}
The tidal radius can be estimated from~(\ref{eq:lambda}) as
\begin{align}\label{eq:rt}
r_t=\bigg(\frac{GM_\bullet}{\lambda_1}\bigg)^{1/3}=\bigg[\frac{M_\bullet}{3M^{\rm dSph}}\bigg(1+\frac{c^{\rm dSph}}{R_\bullet}\bigg)\bigg]^{1/3}R_\bullet,
\end{align}
which recovers the academic point-mass case for $c^{\rm dSph}=0$. Inserting the parameters of the experiments run in this Section into~(\ref{eq:rt}) yields $r_t\simeq 190\pc\approx 10\,r_\epsilon$.

Dotted and solid lines in the upper panel of Fig.~\ref{fig:rotate} show phase-space surfaces that obey the conditions $E=v_x'^2/2+\Phi_\bullet(x',0,0)=0$ and $E_{\rm rot}=v_x'^2/2+\Phi_{\rm eff}(x',0,0)=0$, respectively, with primes denoting quantities measured in a co-rotating frame. By definition, temporary captures (shown in blue) populate a phase-space region where $E<0$ over a wide range of distances. Interestingly, permanent captures are more confined in phase space. In particular, we find that all permanent captures obey $E_{\rm rot}<0$ and can only be found at distances $r<r_t\simeq 10\,r_\epsilon$ from the substructure. Notice also the apparent lack of permanent captures at distances smaller than the thermal critical radius, $|x|\lesssim r_\epsilon$. As discussed below, this is likely due to the finite number of particles in our models.

Lower panel of Fig.~\ref{fig:rotate} shows the energy ($E$) and angular momentum modulus ($L=|\bb r \times \bb v|$) of the particles plotted in the upper panel. For reference, the angular momentum of circular orbits at a given energy is shown with a black-solid line. The first noteworthy result is that permanent and temporary captures occupy different regions of the integral-of-motion space. In particular, it is rare to find permanent orbits with low angular momentum, which suggests that orbits with high-eccentricity do not remain bound to the substructure potential for an indefinite amount of time. In contrast, temporary captures avoid orbits with low eccentricity. This result agrees with the numerical experiments published in Paper I, which show that particles trapped in the potential $\Phi_\bullet$ exhibit a `super-thermal' eccentricity distribution with an excess of particles moving on very eccentric orbits. Notice also that the division between temporary and permanent captures blurs in the energy 'fringe', $|E|\lesssim 0.2\,K$, where $K=3\sigma^2/2$ is the average kinetic energy of field particles. 

Given that captured particles do not conserve $E$ or $L$, as shown in Fig.~\ref{fig:orb}, it is remarkable that temporary and permanent orbits remain locked within their respective regions as they drift in $E$--$L$ space. This empirical result may serve as a starting point for follow up theoretical work to better understand the dynamical mechanisms that lead to gravitational capture.

\begin{figure}
\begin{center}
\includegraphics[width=84mm]{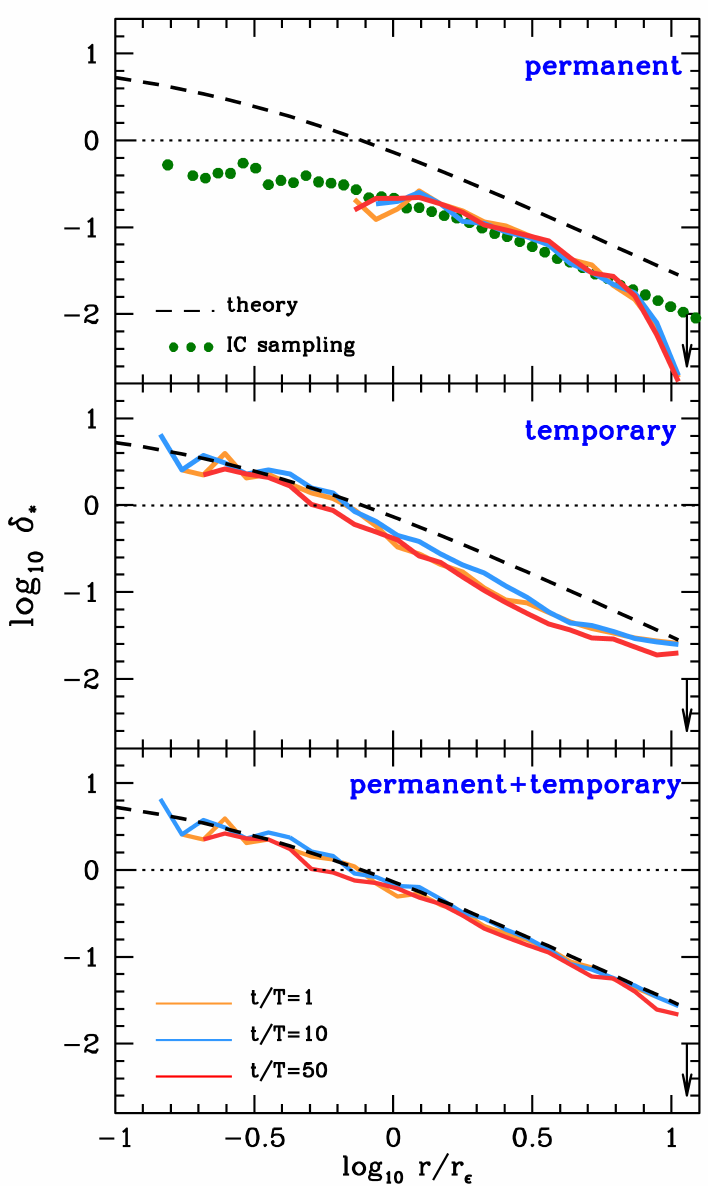}
\end{center}
\caption{Density enhancement ($\delta_\star=n_\star/n$) of particles with negative binding energies $E<0$ as a function of distance from the substructure. Top, middle and bottom panels split bound particles according to whether they are permanent, temporary or any kind of captures (see text). Horizontal dotted lines mark the local density of the field $\delta_\star=1$. Green dots show the density enhancement derived from the equilibrium Initial Conditions (ICs) by placing a substructure Hernquist potential $\Phi_\bullet$ at 100 random locations in the galaxy and identifying particles with $E<0$. Note that lack of permament captures beyond the the tidal radius $r_t$, Equation~(\ref{eq:rt}), marked with vertical arrows. Notice also that the density enhancement profiles do not evolve with time, and that stellar overdensities ($\delta_\star>1$) of field particles mainly originate from temporary captures. Bottom panel shows that the theoretical profiles accurately describe the distribution of all bound particles, that is the superposition of permanent and temporary populations.
}
\label{fig:delta_t}
\end{figure}

Fig.~\ref{fig:delta_t} shows the density enhancement of bound particles, $\delta_\star=n_\star/n$, as a function of distance from the substructure for (i) permanent captures (top panel), (ii) temporary captures (middle panel) and all bound particles (bottom panel) at different snapshots of the simulation. For ease of reference, the background density $\delta_\star=1$ is marked with horizontal dotted lines. As expected, in the static experiments we find a very weak temporal variation of the profiles, suggesting that the distribution of bound particles reaches steady state quickly. 
Focusing first on the top panel, we find that permanent captures do not generate over-densities, i.e. $\delta_\star<1$. We also observe a sharp truncation of the profile at $r\approx 10\,r_\epsilon$, which roughly corresponds to the value estimated from~(\ref{eq:rt}) (marked with vertical arrows). At small distances, $r\lesssim r_\epsilon$, we find no permanent captures. This is likely a numerical artifact resulting from finite sampling. To prove this point, we place a substructure potential $\Phi_\bullet$ at 100 random locations within the galaxy and identify bound particles in the unperturbed models at $t=0$. The averaged density enhancement profile is shown with green dots. Recall that permanent captures are already bound at $t=0$, and given that their spatial distribution shows no temporal variation it is not surprising that these objects follow a similar profile as the one derived from random sampling the Initial Conditions (ICs). However, there are two visible discrepancies: at large radii, $r\gtrsim r_t$, the profile derived from the ICs is not truncated, which indicates that the truncation exhibited by permanent captures originates from dynamical evolution. A second mismatch can be observed at small radii, $r\lesssim r_\epsilon$, where we find that the profile derived from the ICs extends well below that of permanent captures. This is because the IC profile is generated by placing the substructure potential $\Phi_\bullet$ at 100 random locations in the galaxy, which enlarges the statistical sample of bound field stars, allowing us to measure their density at smaller radii. Crucially, the density enhancement converges slowly to unity, $\delta_\star\to 1$ in the limit $r/r_\epsilon\to 0$, which again indicates that the presence of permanent captures does not generate a stellar over-density at the substructure location.

The density enhancement generated by temporary captures is plotted in the middle panel Fig.~\ref{fig:delta_t}. In contrast to  permanent captures, temporary captures generate an over-density ($\delta_\star> 1$) at small distances from the substructure. We also find that the enhancement profile closely follows the theoretical curve~(\ref{eq:delta}) (black-dashed lines) on scales $r\lesssim r_\epsilon$. However, on larger scales $r\gtrsim r_\epsilon$ the density of temporary captures falls off more quickly than predicted by the statistical theory. Beyond the tidal radius, $r\gtrsim r_t$, the profile of temporary captures approaches again the theoretical curve~(\ref{eq:delta}).

Bottom panel show the profiles generated by {\it all} particles with $E<0$ at three different snapshots of the simulation. In agreement with the theoretical predictions from \S\ref{sec:strong}, the theoretical profile~(\ref{eq:delta}) matches the numerical result after we consider the entire population of bound particles without specifying whether captures are temporarily or permanently bound. Notice also that there is no particular feature that marks the location of the tidal radius in the distribution of bound stars, and that Equation~(\ref{eq:delta}) is accurate in the vicinity and beyond the tidal radius, which suggests that the superposition of permanent and temporary orbits conspires to generate a population of bound stars with a constant phase-space density across the volume under observation. We come back to this point below.

\begin{figure}
\begin{center}
\includegraphics[width=72mm]{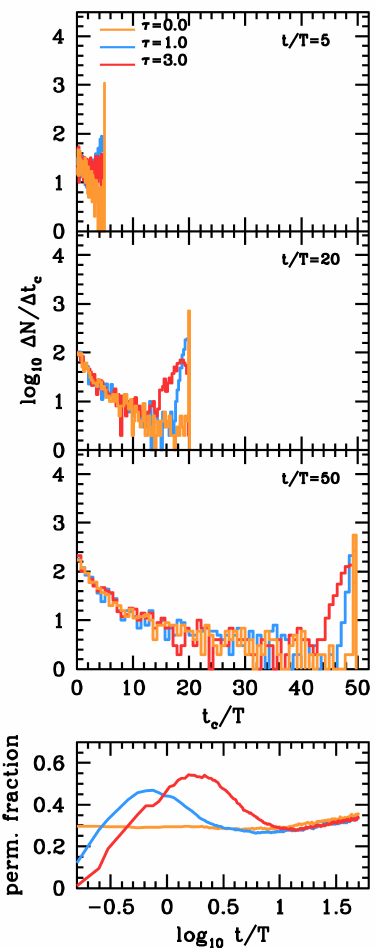}
\end{center}
\caption{{\it Upper panel:} Distribution of capture times ($t_c$) defined as the time since a particle becomes bound (energy flips from positive to negative) to the current simulation time ($t$). In substructure models with different growth rates ($\tau$), we find a substantial number of particles captured at early times ($t_{\rm bound}\lesssim 2\tau$) that remain bound for the rest of the integration, the so-called `permanent' captures, which materialize as a peak in the distribution at long capture times $t_c\sim t$, which shifts towards the right of this plot as the integration time increases. 
{\it Lower panel:} Fraction of permanent captures within a volume element $r_{\rm out}=190\pc$ as a function of integration time. The fraction of permanent captures is roughly constant in the static case ($\tau=0$), and peaks around $t\sim \tau$ for substructures with a time-dependent mass~(\ref{eq:Mt}). At time-scales $t\gg \tau$, the three models show the same fraction of permanent particles independently of $\tau$.
}
\label{fig:time_ptt_taum_4}
\end{figure}

\subsection{Growing substructures on a circular orbit}\label{sec:growth}
In the previous Section, a static substructure potential $\Phi_\bullet$ is instantly injected in a sea of field particles in dynamical equilibrium. This experimental set up is mathematically convenient because it removes any explicit time dependence from the analytical estimates, but it has limited physical applications.

In this Section, we inspect a physically-motivated scenario in which a substructure has a mass that {\it grows} as it moves along a fixed orbit across a sea of field particles (see discussion in \S\ref{sec:dis}). 
To this aim, 
we re-run the 'compact' $N$-body models presented in \S\ref{sec:ptt} adopting a time-dependent substructure mass
\begin{align}\label{eq:Mt}
M_\bullet(t)=M_\bullet \,[1-\exp(-t/\tau)],
\end{align}
with $M_\bullet=10^6\msol$ and a fixed scale radius $c_\bullet=3.78\pc$. Here, $\tau$ is a parameter that controls the mass growth rate, $\tau=|\dot M_\bullet/M_\bullet|^{-1}_{t=0}$.
The substructure potential $\Phi_\bullet$ vanishes in the limit $t\to 0$ for $\tau>0$, which allows us to easily modify the number of immediate captures by choosing different values of the time-scale $\tau$. By construction, in the limit $t\gg \tau$ the time-dependent model approach the static case explored in \S\ref{sec:ptt}.

Fig.~\ref{fig:time_ptt_taum_4} shows the distribution of capture times, $t_c=t-t_{\rm bound}$, where $t_{\rm bound}$ is the integration time at which a particle becomes bound, in models with three different mass-growth rates, $\tau/T=0, 1$ and $3$. Orange lines correspond to the same distribution plotted in Fig.~\ref{fig:time} for static models ($\tau=0)$. Comparison between the panels from top to bottom shows the emergence of prominent spike of particles with $t_c\approx t$, which is the telltale of 
the build-up of a permanent captures trapped in the time-evolving substructure potential, $\Phi_\bullet(t)$. 
Focusing on the models with a slow growth rate ($\tau=3\,T$, red lines) we find that the spike is absent early on during the simulation ($t/T=5$, top panel), and that it only starts to arise at $t/T\gtrsim 20$ (second panel from the top).
At the end of the simulation at $t=50\,T$ (third panel from the top), we find that all three models show a clear excess of captures with long capture times, $t_c\sim t$, with the peak of the distribution broadening up for models with slow growth rates ($\tau>0$). This suggests that while in a static potential ($\tau=0$) permanent captures are trapped immediately at the start of the experiment, $t_{\rm bound}=0$, this is not the case in substructures with a mass growth~(\ref{eq:Mt}), which capture particles onto permanent orbits over an extended period of time, $t_{\rm bound}\lesssim 2\tau$.

\begin{figure}
\begin{center}
\includegraphics[width=84mm]{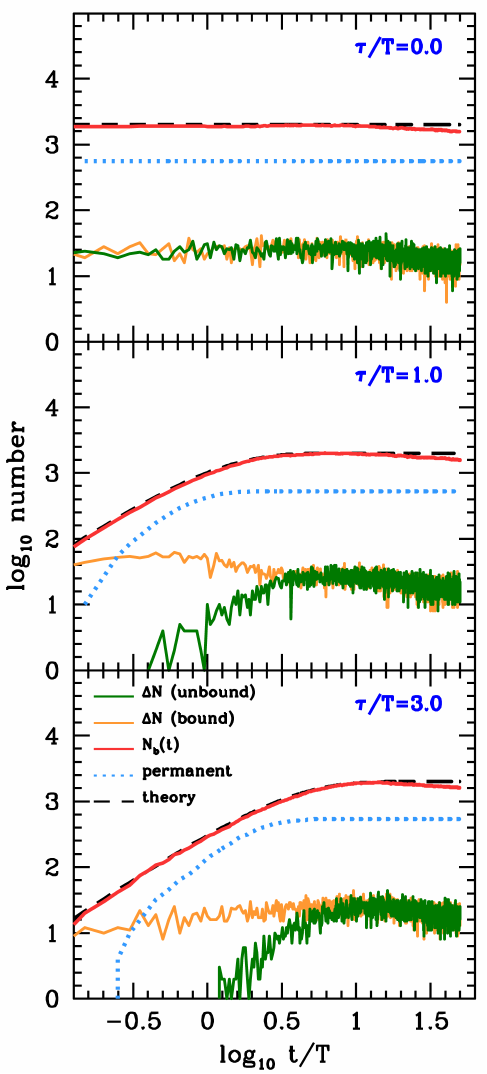}
\end{center}
\caption{{\it Upper panel:} Distribution of capture times ($t_c$) defined as the time since a particle becomes bound (energy flips from positive to negative) to the current simulation time ($t$). Notice the presence of a substantial number of particles that become bound at $t\approx 0$, which result in a a strong peak in the distribution at $t_c\approx t$ shifting towards the right as the integration time increases.
{\it Lower panel:} Permanent-to-temporary ratio (or ptt in short) of particles captured within a volume element $r_{\rm out}=100\pc$ as a function of integration time. Permanent captures are defined as particles with a capture time $t_c\approx t$, i.e. they remain bound during the whole integration time. In contrast, temporary captures only remain bound for a fraction of the integration time, i.e. $t_c\lesssim t$. Notice that the ptt ration becomes approximately constant at $t\gtrsim 5,T$, where $T$ is Smoluchowski's time scale.
}
\label{fig:numt_3}
\end{figure}

Unfortunately, the theoretical conditions that determine whether a field particle becomes permanently or temporarily bound to a time-dependent potential $\Phi_\bullet(t)$ are more difficult to study than in the static case, because the energy measured in a co-rotating frame ($E_{\rm rot})$ is not a conserved quantity in time-dependent systems. 
The results shown in Fig.~\ref{fig:time_ptt_taum_4} call for a heuristic classification of orbits according the the amount of time that they remain bound. In particular, in what follows we identify {\it permanent} captures as particles that become bound early on, $t_{\rm bound}\le 2\tau$, and whose energy $E$ remains negative uninterruptedly until the end of the simulation\footnote{Notice that some temporary captures may also remain bound for relatively long times. However, Fig.~\ref{fig:time} suggests that the fraction of temporary captures that form within $t_{\rm bound}\le 2 \tau$ is $\lesssim 10\%$}. These are the particles that populate the spikes of the distributions plotted in Fig.~\ref{fig:time}.
With this empirical definition at hand, it is straightforward to measure the fraction of bound particles that are captured on permanent orbits as a function of time, which is shown in the bottom panel. The first noteworthy result is that the fraction of permanent captures is approximately constant in the static model\footnote{The potential $\Phi_\bullet$ moving across the galaxy heats up the population of field particles. This translates into a capture rate that decreases with time (see also Fig.~\ref{fig:numt_3}), and to a slightly rising fraction of permanent captures at $t\gtrsim 10\,T$.}, $f_{\rm perm}=N_{\rm perm}/(N_{\rm perm}+N_{\rm temp})=N_{\rm perm}/N_b\simeq 0.35$
As expected, increasing the time-scale $\tau$ systematically decreases the number of permanent captures at $t=0$. In particular, the model with the slowest growth ($\tau=3\,T$, red line) contains no permanent captures initially. As time proceeds, the number of permanent captures grows, peaking at $t\approx \tau$. Once the substructure stops growing, the fraction of permanent captures approaches the static value at $t\gg \tau$, independently of the choice of $\tau$.
This suggests that the exact manner in which $\Phi_\bullet$ is inserted into the field does not affect the final distribution of bound particles, insofar as the captured population has enough time to reach steady state. 

Fig.~\ref{fig:numt_3} plots the number of bound particles located within a fixed spherical radius $r_{\rm out}=190\pc$ from the centre of $\Phi_\bullet(t)$ as a function of time (red lines) for three different values of the time-scale $\tau$. Recall that in the static case, this choice of $r_{\rm out}$ matches the local tidal radius~(\ref{eq:rt}).
We find that the total number of particles with $E=v^2/2+\Phi_\bullet(t)<0$ within the volume ($N_b$) grows in proportion to the substructure mass. The experimental values are in excellent agreement with the values derived from Equation~(\ref{eq:Nb}) using a time-dependent mass~(\ref{eq:Mt}) (black-dashed lines). 

 Fig.~\ref{fig:numt_3} also illustrates the process by which a population of captured particles reaches steady state. Orange and green lines respectively show the number of particles in the volume $V$ that become bound ($E$ flips from positive to negative) and unbound (either $E$ flips from negative to positive, or the particle leaves the volume under observation) within a time interval $\Delta t=0.1\,T$. In models where $\Phi_\bullet$ is introduced suddenly ($\tau=0$), we find that these two values become approximately equal after a few snapshots. At this point, steady state has been reached. In models where the potential $\Phi_\bullet$ grows slowly ($\tau>T$) we observe two distinct regimes: (i) at $t\lesssim \tau$, particles are captured at a rate that overcomes the number of unbinding events, which leads to a net growth of the population of captured particles. (ii) On longer time-scales, $t\gtrsim \tau$, the number of bound particles reaches a steady state, wherein the rate of capture becomes roughly equal to the loss rate, $\Delta N_{\rm bound}\approx\Delta N_{\rm unbound}$. From this time on, all captures are temporarily bound, which leads to no net variation in the average number of stars with $E<0$. It is worth to highlight that in all models the capture rate and the loss rate conspire to yield an average number of bound particles within the volume $V$ that is very close to the value derived from Equation~(\ref{eq:Nb}) with a time-dependent mass~(\ref{eq:Mt}) {\it at all times}. 
 
Blue-dotted lines in Fig.~\ref{fig:numt_3} show the number of permanent captures as a function of time. Notice that all particles trapped onto permanent orbits are captured early on, $t\lesssim 2\tau$, and that at the end of the simulation $t_f=50\,T$ all models contain a similar number of permanent captures independently of the choice of $\tau$. At early times, the build up of permanently-bound orbits leads to a capture rate that exceeds the loss rate, $\Delta N_{\rm bound}>\Delta N_{\rm unbound}$, and thus to a net growth of the average number of bound particles, $N_b(t)$. This may suggest that permanent captures result from an {\it impulsive} response of the galactic field to the growth of the substructure potential on short time-scales $|\dot \Phi_\bullet/\Phi_\bullet|^{-1}=|\dot M_\bullet/M_\bullet|^{-1}_{t=0}=\tau\sim T$. Further theoretical work is need to understand how this capture mechanism works.

\begin{figure}
\begin{center}
\includegraphics[width=78mm]{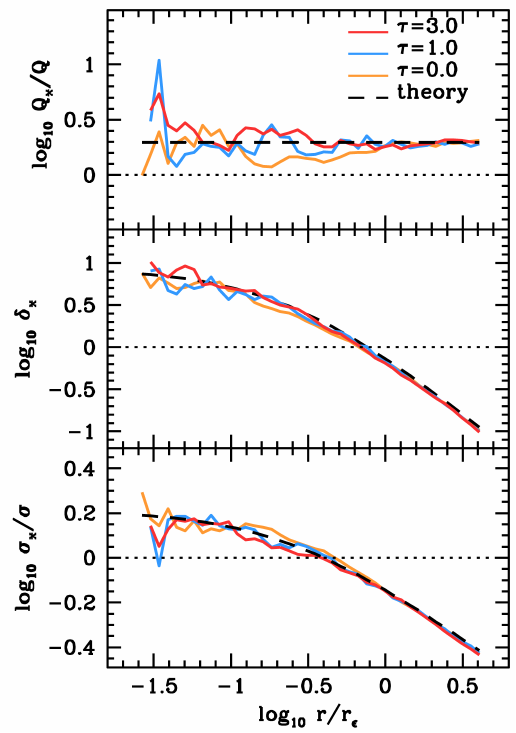}
\end{center}
\caption{{\it Upper panel}: Average phase-space density $Q_\star=n_\star/\sigma^3_\star=n\,\delta_\star/\sigma^3_\star$ as a function of distance from the potential $\Phi_\bullet$ measured from the models shown in Fig.~\ref{fig:numt_3}. Distances are measured in units of the thermal critical radius $r_\epsilon$, Equation~(\ref{eq:r_eps}). Note that $Q_\star$ remains approximately constant within the volume element, and that its value agrees well with the theoretical expectation~(\ref{eq:Qstar}) (black-dashed line). {\it Middle panel}: Density enhancement profile $\delta_\star(r)=n_\star(r)/n$. The black-dashed line shows the theoretical curve~(\ref{eq:delta}). For reference, we mark with a dotted line the background density value $\delta_\star=1$. {\it Lower panel}: Velocity dispersion profiles of the models shown above normalized by the local velocity dispersion of the field, $\sigma_\star(r)/\sigma$. Black-dashed line shows the theoretical profile~(\ref{eq:vdisp}).  }
\label{fig:prof_3_taum}
\end{figure}

Fig.~\ref{fig:prof_3_taum} shows that the final distribution of bound particles around the potential $\Phi_\bullet$ at $t=t_f$ is not sensitive to the rate of growth of the substructure potential ($\tau$). This result holds insofar as the simulation time is longer than the time-scale on which the potential grows, i.e. $t_f\gg \tau$. 
Let us focus first on the upper panel, which plots the average phase-space density of captured particles as a function of distance to the substructure, $Q_\star(r)\equiv n_\star(r)/\sigma^3_\star(r)$. As expected, we find that $Q_\star$ is approximately constant across the volume $V$, and that the measured value of $Q_\star$ approximately matches Equation~(\ref{eq:Qstar}) (marked with a black-dashed line) independently of the choice of $\tau$. 
Section~\ref{sec:strong} predicts that bound particles with a constant phase-space density in steady state follow the density profile~(\ref{eq:delta}) (black-dashed line). The middle panel shows that this expectation is largely correct. Regardless of whether $\Phi_\bullet$ is introduced gradually ($\tau>0)$ or suddenly ($\tau=0$), we find that the number density of captured particles scales as $\delta_\star\sim|\Phi_\bullet|^{3/2}$. Similarly, the lower panel of Fig.~\ref{fig:prof_3_taum} shows that the velocity dispersion profile goes as $\sigma_\star\sim|\Phi_\bullet|^{1/2}$, which in the case of a Hernquist substructure potential $\Phi_\bullet$ leads to the velocity dispersion profile~(\ref{eq:vdisp}) plotted with a black-dashed line.

\begin{figure}
\begin{center}
\includegraphics[width=86mm]{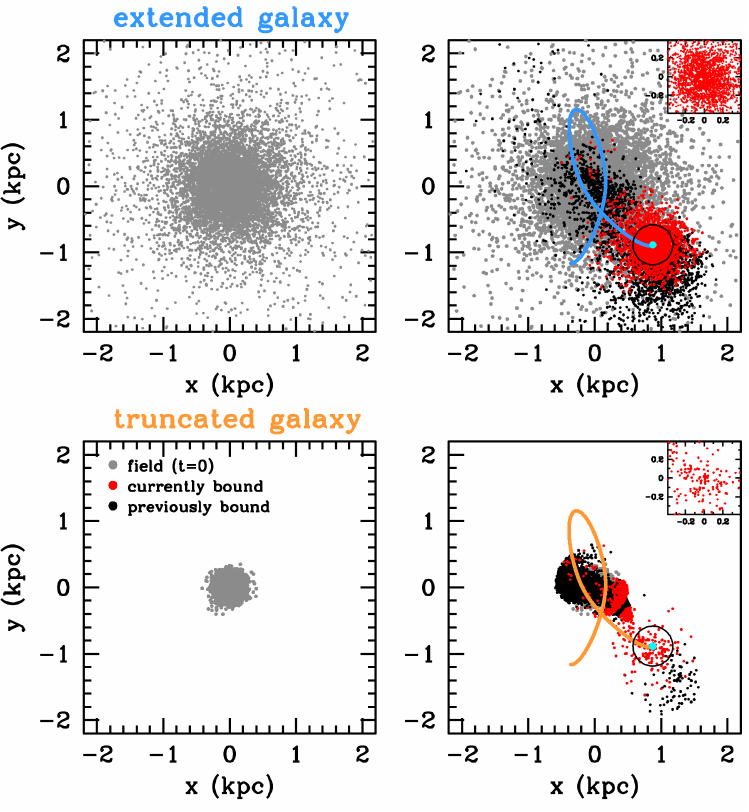}
\end{center}
\caption{
{\it Left panels:} Initial distribution of field particles (grey dots) in dynamical equilibrium within a Dehnen potential with a total mass $M^{\rm dSph}=3\times 10^9\msol$, a scale radius $c^{\rm dSph}=2\kpc$ and a central density slope $\gamma^{\rm dSph}=1$.  
{\it Right panels:} Distribution of particles that become bound to a Hernquist potential with $M_\bullet=3\times 10^7\msol$ and a compactness $\kappa=+0.9$ (see text). Particles that become captured during the integration time and have $E<0$ ($E>0$) at the final snapshot ($t=t_f$) are highlighted with red (black) dots. The trajectory of the substructure across the galaxy is shown with a solid line. Black circles mark a volume size $r_{\rm out}=300\pc$ around the substructure potential. Inlets in the upper-right corner zooms in at the substructure location (marked with a cyan dot for ease of reference). Distances are given in $\kpc$. Notice that particles captured from a truncated galaxy exhibit a flattened shape roughly oriented along the orbital motion.
}
\label{fig:xyz}
\end{figure}

\subsection{Accreted substructures on an eccentric orbit}\label{sec:accretion}
In the previous Sections, we carry out experiments where a substructure is injected on a circular orbit either suddenly ($\tau=0$), fast ($\tau=T$), or slowly ($\tau>T)$ into a pre-existing field of field particles in dynamical equilibrium. This experiminental setup invariably leads to a population of field particles that are captured onto permanent orbits around the potential $\Phi_\bullet(t)$. 
In this Section, we present models where a static substructure potential is initially placed in a region of the galaxy devoid of field stars on an eccentric orbit. By design, these experiments contain no immediate or permanent captures. 

The numerical set-up is slightly different from previous Sections.
Here, we consider a dwarf galaxy sourcing a cuspy Dehnen (1993) potential ($\gamma=1$) with a mass $M^{\rm dSph}=3\times 10^9\msol$ and a scale radius $c^{\rm dSph}=2\kpc$.
For illustration, we show two experiments where the substructure is inserted in the galactic potential at apocentre, $R_{\rm apo}=1380\pc$. The initial velocity is set to $V_\bullet=7.9\kms$, which leads to a pericentre of $R_{\rm peri}=165\pc$. Thus, this orbit penetrates the inner-most regions of the galaxy. We then follow the motion of the substructure for two dynamical times measured at apocentre, $\Omega^{-1}(R_{\rm apo})=R_{\rm apo}/V_c(R_{\rm apo})=34\myr$, where $V_c(R_{\rm apo})=39.4\kms$ is the circular velocity. Before showing the outcome of this experiment, it should be pointed out that we have explored several combinations of apo- and peri-centres and number of orbital periods, finding similar results as discussed below.


We consider two populations of tracer particles in equilibrium within the galactic potential (see \S\ref{sec:nbody} for details).
The first population follows a `truncated' density profile $(\alpha_f,\beta_f,\gamma_f)=(2,30, 0.2)$ with a scale radius $R_0=460\pc$, which falls off very steeply beyond $R\gtrsim 600\pc$. The luminosity-averaged velocity dispersion of these particles is $\langle v^2\rangle^{1/2}/3=11.7\kms$. The second population follows an `extended' Plummer-like profile with $(\alpha_f,\beta_f,\gamma_f)=(2,5, 0.1)$ and the same scale radius, $R_0=460\pc$, which leads to a slightly higher luminosity-averaged velocity dispersion $\langle v^2\rangle^{1/2}/3=19.9\kms$. 

The substructure sources a static Hernquist potential with a mass $M_\bullet=3\times 10^7\msol$.
To set the scale radii of the models we use the thermal critical radius at $t=0$ as a reference. For illustration, we set the compactness to $\kappa=+0.9$ in both galaxy models, which leads to a scale radius $c_\bullet=(1-\kappa)r_\epsilon=0.1\,r_\epsilon$. To find the thermal critical radii of the two models we insert $\sigma=\langle v^2\rangle^{1/2}/3$ and $V_\bullet$ in Equation~(\ref{eq:r_eps}), which yields $r_\epsilon=390\pc$ and $980\pc$ for the extended and truncated galaxy models, respectively.


Left panels of Fig.~\ref{fig:xyz} show the projected locations of tracer particles with extended (top panel) and truncated (bottom panel) profiles in equilibrium ($t=0$). In the right panels, we overplot the trajectory of the substructure as it falls into the galaxy on an eccentric orbit (solid lines), and mark its final location with a cyan dot. Red dots show particles with $E<0$ at the final snapshot $(t_f)$. Black dots correspond to particles that have been captured for a finite amount of time before being lost to galactic tides. 

As expected, zooming in at the substructure location we find that particles with $E<0$ generate a local overdensity of field particles. Interestingly, their observed spatial distribution depends strongly on the initial profile of the field. In the extended galaxy model (top panel), the over-density of bound particles has a close-to-spherical shape and a relatively large size. In contrast, the over-density in the bottom panel has a smaller size and a flattened shape, which appears to be roughly oriented along the orbital motion.

\begin{figure}
\begin{center}
\includegraphics[width=80mm]{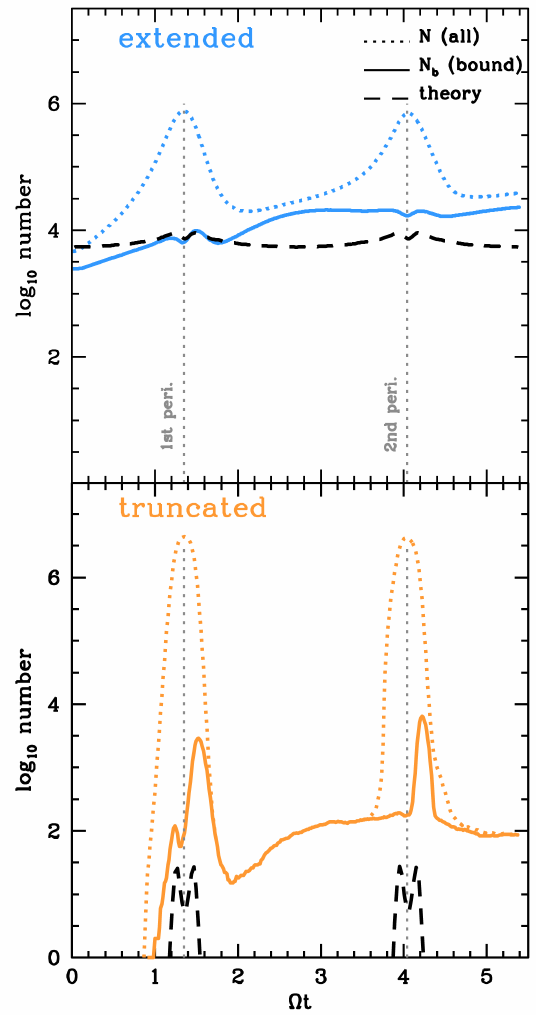}
\end{center}
\caption{Number of bound particles located within a volume size $r_{\rm out}=0.3\kpc$ as a function of time for the models plotted in Fig.~\ref{fig:xyz}. Dotted lines show the total (bound plus unbound) number of field particles within the volume, $r<r_{\rm out}$. Black-dashed lines show the theoretical prediction from Equation~(\ref{eq:Nb}), with a time-dependent number density and velocity dispersion of field particles measured at the location of the substructure, $n(t)=n[R(t)]$ and $\sigma(t)=\sigma[R(t)]$. Pericentric passages are marked with vertical dotted lines.
}
\label{fig:numt2}
\end{figure}

The rate at which field particles are captured in these two models is also strikingly different.
Solid lines in Fig.~\ref{fig:numt2} shows the number of bound particles within a volume size $r_{\rm out}=300\pc$ (marked with black circles in Fig.~\ref{fig:xyz} for reference) as a function time. In the top panel we can see that the substructure orbiting in the extended galaxy contains $N_b\sim 3\times 10^3$ at $t=0$. By definition, these are {\it immediate} captures. As the substructure falls into the inner regions of the galaxy, the total number of field particles within the volume (dotted lines) increases systematically until pericentre is reached at $\Omega\,t\simeq 1.4$. 
After the pericentre passage, the number of bound particles (solid lines) grows slightly, reaching a plateau at apocentre, $\Omega\,t\simeq 2.8$.
This situation repeats throughtout the next orbital revolution.
At the last snapshot of the simulation, the number of bound particles is $N_b(t_f)\sim 4\times 10^4$, which is a factor $\sim 10$ increases with respect to the value at $t=0$. 
Remarkably, the theoretical expectation given by Equation~(\ref{eq:Nb}) (black-dashed line) barely changes as a function of orbital phase in spite of the large variation of galactocentric distances and velocities. This reflects the counter balance between the variation of the mean phase-space density of field particles along the orbit, $Q(R)=n(R)/\sigma^3(R)$, and the exponential suppression of captures at high velocities, $\xi=\exp[-V^2_\bullet/(2\sigma^2(R))]$. As the substructure plunges into the inner regions of the galaxy the orbital velocity increases, making capture less efficient. Simultaneously, the number of field particles around the substructure grows, with both effects nearly balancing out in Equation~(\ref{eq:Nb}). The reverse process takes place from pericentre to apocentre.

The substructure model orbiting within a truncated galaxy experiences a very different evolution. In this case, bottom panel of Fig.~\ref{fig:numt2} shows that capture does not occur until the substructure is close to its first orbital pericentre. As expected, the number of immediate captures is zero by construction. After each pericentric passage, we observe a very rapid increase in the population of bound particles, whose size can grow by $\sim 3$ orders of magnitude. However, a large fraction of these particles are quickly lost to galactic tides. At first apogalacticon, the number of bound particles has stabilized at $N_b\sim 10^2$. As it reaches its second pericentre, it grows again up to $N_b\sim 5\times 10^3$, falling back to $N_b(t_f)\sim 10^2$ at its second apocentre. 
Comparison with the theoretical expectation (black-dashed lines) shows that Equation~(\ref{eq:Nb}) cannot accurately describe this behaviour. There are two main reasons for the mismatch. In the first place, the statistical theory derived in \S\ref{sec:number} relies on the local approximation, which assumes that the properties of the galactic field do not vary strongly as a function of distance from the substructure. This is clearly not the case in this particular galaxy model, which is strongly truncated beyond $R\gtrsim 600\pc$. Furthermore, it assumes that the properties of the field are time-invariant, and that the population of captured particles is in steady state at all times. None of these conditions apply to the the model shown in the bottom panel. The latter approximation is particularly poor after the first pericentric passage. Here, Equation~(\ref{eq:Nb}) predicts $N_b=0$ because the density of field particles at this location is $n(R_{\rm apo})=0$ (see bottom-left panel of Fig.~\ref{fig:xyz}). In contrast, our numerical experiment shows that the substructure has captured $N_b\sim 10^3$ particles after the first orbital revolution. 
This mismatch arises again at the second apocentre, which corresponds to the final snapshot of the simulation. Here, all particles within the volume under observation (dotted line) are bound to the substructure (solid line). It is worth stressing that region of the galaxy was empty from field particles at $t=0$. Looking the bottom-right panel of Fig.~\ref{fig:xyz}, we find field particles beyond the original truncation radius of the galaxy are either still bound to the substructure (red dots), or were `scooped' from the inner regions and released back to the galaxy by tidal forces (black dots).

\begin{figure}
\begin{center}
\includegraphics[width=84mm]{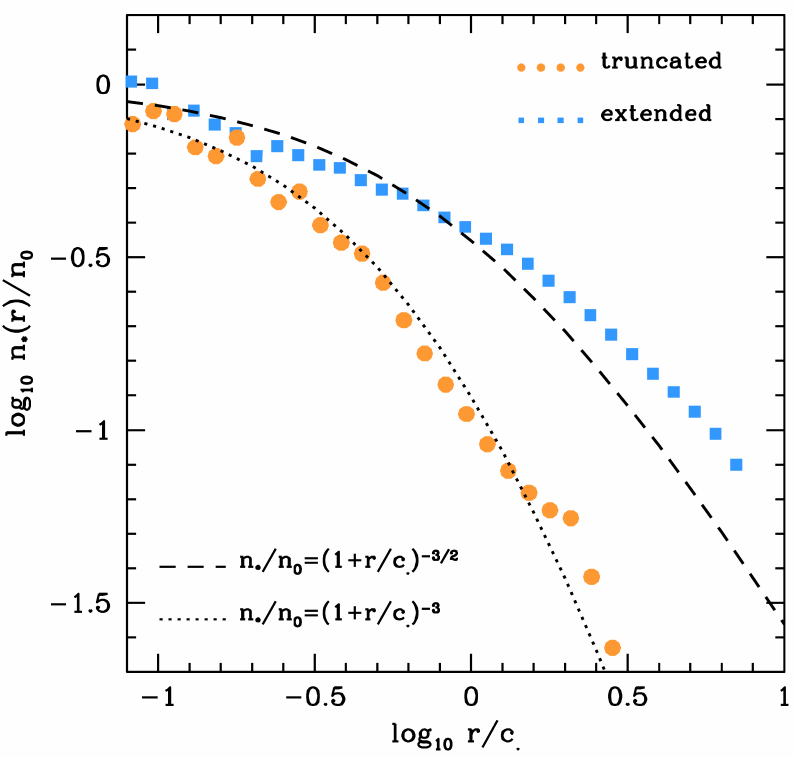}
\end{center}
\caption{Number density profile of field particles bound to the substructure models shown in Fig.~\ref{fig:xyz}. Distances are measured relative to the final substructure location and given in units of the substructure scale radius ($c_\bullet$). The profiles are averaged over 500 realizations of the model to reduce statistical noise (see text). Notice that the model associated with an extended galaxy model roughly follows the theoretical profile~(\ref{eq:delta}). In contrast, substructures evolving in a truncated galaxy generate a local overdensity that falls off steeply at large distances $\delta_\star \sim r^{-3}$ at $r\gg c_\bullet$.
}
\label{fig:delta_trun}
\end{figure}

Fig.~\ref{fig:delta_trun} shows that distribution of bound particles in the extended and truncated galaxies plotted in Fig.~\ref{fig:xyz} follow very different profiles. For ease of comparison, we normalize the profiles to its central value, $n_0=n_\star(r=0)$, and measure distances in units of the substructure scale radius ($c_\bullet$). Given the relative small number of particles with $E<0$ at the final snapshot, we reduce statistical noise by generating 200 and 600 realizations of the extended and truncated galaxy models, respectively. This is done placing substructures at random positions over the surface of a sphere with a radius $R_{\rm apo}=1380\pc$ and a tangential velocity $V_\bullet=7.9\kms$ at $t=0$, such that all individual realizations have a common pericentre of $R_{\rm peri}=165\pc$.

Particles captured by the substructure model orbiting in an extended galaxy exhibit a profile shape that roughly scales as $(1+r/c_\bullet)^{-3/2}$, which matches Equation~(\ref{eq:delta}). This is a remarkable result considering the strong strong temporal evolution of the field along the substructure orbit, and that the statistical theory is derived for static systems. However, the normalization of the profile is not so well matched by the analytical formula. In particular, Fig.~\ref{fig:numt2} shows that Equation~(\ref{eq:delta}) tends to underpredict the number of field particles bound to the substructure at all points of the orbit. 

This mismatch is considerably stronger in substructures orbiting a in a truncated galaxy. In these objects, captured field particles show a profile that falls off more steeply than predicted by Equation~(\ref{eq:delta}) at large distances, $r\gtrsim c_\bullet$, whereas at small distances, $r\ll c_\bullet$, it converges to the profile of the extended galaxy model. The different behaviour may be traced back to the lack of permanent captures in the truncated galaxy models, which tend to smooth out the overall density enhancement profile (see Fig.~\ref{fig:delta_t}).

In general, the numerical experiments in this Section indicate that the distribution of field stars captured by substructures moving on eccentric orbits vary as a function of time in different ways depending on whether these objects are form in a pre-existing sea of field particles, or are accreted from outside the galaxy.

\begin{figure*}
\begin{center}
\includegraphics[width=170mm]{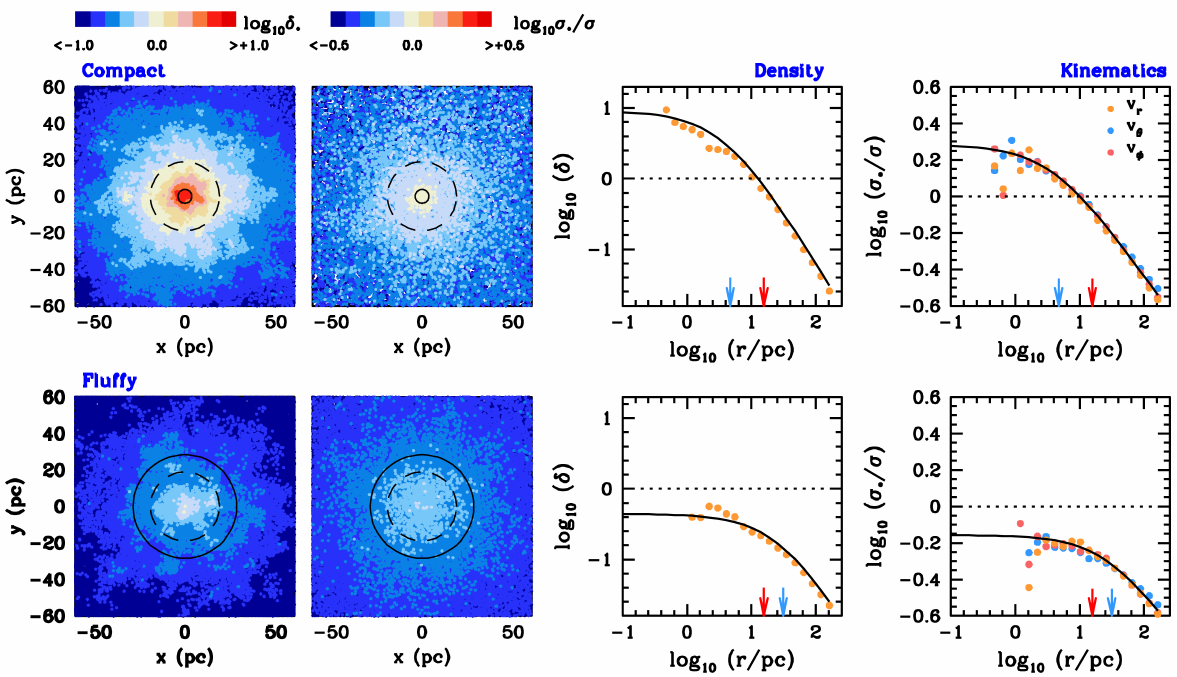}
\end{center}
\caption{{\it Upper left panels:} Locations projected onto the orbital plane of stellar tracer-particles temporarily-bound to a substructure with a mass $M_{\bullet}=10^6\msol$ moving on a circular orbit at a fixed distance $R=0.5\kpc$ from the centre of the Fornax dSph (see \S\ref{sec:nbody}). These models adopt two different scale-radii (marked with black-solid circles for reference): 'compact' ($\kappa=+0.8$, or $c_\bullet=0.2\,r_\epsilon$ (upper panels) and `fluffy' ($\kappa=-0.5$, or $c_\bullet=1.5\,r_\epsilon$ ( lower panels), where $r_\epsilon=18.9\pc$ is the thermal critical radius measured from~(\ref{eq:r_eps}) (black-dashed circle). {\it Middle left panels:} Velocity dispersion map for the models shown in the left panels. 
{\it Middle right panels:} Density enhancement as a function of distance to the substructure centre. Red and blue arrows mark the location of the thermal critical radius and the scale length of the substructure, respectively. The theoretical profile predicted by Equation~(\ref{eq:delta}) is shown with a black line. {\it Right panels:} Velocity dispersion profiles in spherical coordinates of the above models. Black lines show the profiles predicted by Equation~(\ref{eq:vdisp}). Captured particles have isotropic velocities on account of their chaotic motions in the dwarf potential (see text). }
\label{fig:map_v}
\end{figure*}

\subsection{Compact vs. fluffy substructures}\label{sec:fluffy}
The statistical theory outlined Section~\ref{sec:number} predicts that only dark subhaloes that are sufficiently {\it compact} ($\kappa>0$) can become visible as localized overdensities of field stars. 
To illustrate this result, this Section presents similar experiments as shown in \S\ref{sec:ptt} with positive and negative compactness values.

Fig.~\ref{fig:map_v} shows the projection onto the orbital plane of particles bound to a Hernquist substructure potential with a mass $M_\bullet=10^6\msol$ moving on a circular orbit $R_\bullet=0.5\kpc$ in the Fornax dSph-like galaxy model introduced in \S\ref{sec:nbody}. Recall that the tidal radius given by Equation~(\ref{eq:rt}) is $r_t\approx 190\pc$. 
We consider two models with $\kappa=+0.8$ (upper panels) and $\kappa=-0.5$ (lower panels), with the negative sign of $\kappa$ implying that the number density of captured field particle does not exceed the background ($\delta_\star<1)$ at any distance from the substructure. 
Application of Equation~(\ref{eq:r_eps}) yields a thermal critical radius $r_\epsilon=18.9\pc$, which translates into scale-radii $c_\bullet=3.78\pc$ and $c_\bullet=28.3\pc$ for the `compact' and `fluffy' models, respectively. Notice that the time-scale needed to cross the over-density size by a random field particle is much shorter than the orbital time around the dSph galaxy potential,
$T_\epsilon=r_\epsilon/\sigma^{\rm dSph}=18.9\pc/(12.2\kms)\approx 1.5\myr$. As a result, the population of captured particles reaches steady state quickly.
We choose a volume size around the substructure equal to its tidal radius~(\ref{eq:rt}), $r_{\rm out}=r_t=190\pc$. 
We run both models from $t=0$ until $t=50\,T$.

Fig.~\ref{fig:map_v} highlights the impact of the substructure {\it compactness} ($\kappa=1-c_\bullet/r_\epsilon$) on the spatial and kinematic distribution of stars with $E<0$. As predicted in \S\ref{sec:number}, at a fixed mass only compact ($\kappa>0$) objects lead to the formation of over-dense regions of trapped stars with a velocity dispersion that is systematically hotter than in fluffy ($\kappa<0$) counterparts.
These differences can be better quantified in the middle-right and right-most panels, which compare the density enhancement~(\ref{eq:delta}) and velocity dispersion~(\ref{eq:vdisp}) profiles derived analytically (black-solid lines) against the numerical values (orange dots). 
For reference, Fig.~\ref{fig:map_v} marks with vertical arrows the location of $c_\bullet$ (in blue) and $r_\epsilon$ (in red), respectively. At large distances $r\gg c_\bullet$, we find that the density enhancement approaches the Keplerian behaviour $\delta_\star\sim r^{-3/2}$ and the velocity dispersion drops off as $\sigma_\star\sim r^{-1/2}$, whereas at small distances $r\ll c_\bullet$ both profiles become roughly constant. However, while in the compact model ($\kappa=+0.8$) trapped stars reach a relatively high density, $\delta_{\star,0}=\delta_\star(r=0)\approx 9.0$, and a hot velocity dispersion, $\sigma_{\star,0}/\sigma=\sigma_{\star}(r=0)/\sigma\simeq 1.9$, stars captured by the fluffy counterpart with a negative compactness ($\kappa=-0.5$) show subdominant densities, $\delta_{\star,0}=0.44$, and a cold dispersion, $\sigma_{\star,0}/\sigma\simeq 0.69$, with respect to the field. We discuss the relevance of this result for the detection of dark substructures in \S\ref{sec:dis}.

An important property of the models plotted in Fig.~\ref{fig:map_v} is that they appear approximately spherical and show no signature of tidal tails. This is in stark contrast with the substructures moving on eccentric orbits plotted in Fig.~\ref{fig:xyz}, which are elongated and exhibit prominent tidal tails roughly aligned with the orbit of the substructure. This emphasizes the need to model the motion of the substructure in the host potential and the distribution of field particles simultaneously, as discussed in the following Section.

\section{Discussion}\label{sec:dis}

\subsection{Anomalous stellar clusters in dSphs}\label{sec:data}
Several Milky Way dSphs contain stellar substructures that may be good candidates for being composed of field stars captured by dark substructures, i.e. they have a large size for their luminosity, their stellar populations are indistinguishable from those of the host galaxy, and they exhibit DM-dominated mass-to-light ratios.

For example, the Fornax dSph is known to contain 5 massive ($\sim10^5\,{\rm M}_\odot$) globular clusters (e.g. Larsen et al. 2012).
There has been a long-standing debate in the literature about a possible sixth cluster, named Fornax 6 (F6), which was first noted by Shapley (1939). While early studies debated whether it was composed of stars or background galaxies (e.g. Stetson et al. 1998), recent ground-based and {\it Gaia} data show that F6 is clearly an overdensity of stars (Wang et al. 2019).
F6 has total stellar mass of $M_\star^{F6}=(7.2\pm 2.2)\times 10^3 \msol$, and a half-light radius of $r_h^{F6}=11\pm 1.4\pc$ (Wang et al. 2019). 
Crucially, the metallicity $\mathrm{[Fe/H]} = -0.71 \pm 0.05$ and age $\sim 2\gyr$ of F6 are very similar to the average metallicity and age of Fornax's MR stars (Pace et al. 2021). This is in stark contrast with the other 5 globular clusters, which have much lower metallicities that range between $-2.5< \mathrm{[Fe/H]} < -1.4$. With this data at hand F6 members are therefore indistinguishable from MR stars in the Fornax dSph. Importantly, the velocity dispersion of F6 is unexpectedly high, $\sigma^{F6}= 5.6 \pm 2.0\kms$, which translates into an `anomalous' mass-to-light ratio of $15 < M/L < 258$ when virial equilibrium is assumed (Pace et al. 2021). The inflated mass-to-light ratio has been interpreted as a result of on-going tidal disruption, although no tidal tails originating from the cluster have been found. 

Alternatively, the unusual properties of F6 can be explained if this stellar system is made of field stars temporarily captured by a dark substructure orbiting in the Fornax dSph potential. According to Fig.~\ref{fig:kml} a stellar over-density with $N_\star\sim 10^4$ members and a mass-to-light ratio of $M/L\sim 100$ would require a minimum subhalo mass of $M_\bullet\sim 10^6\msol$, which lies significantly below the minimum subhalo mass that can trigger star formation, and is consistent with our working assumption that this object would be `dark'. To be visible as an over-density, it must be compact enough, i.e. $\kappa>0$ or $c_\bullet\lesssim r_\epsilon\sim 20\pc$ (for illustrative examples, see the numerical models presented \S\ref{sec:nbody}). Interestingly, this size is smaller than the average peak-velocity radius of {\it field} CDM haloes of comparable masses, i.e. $r_{\rm max}\sim 100\pc$ for $M_{200}\sim 10^6 \msol$ at redshift $z=0$. We discuss the implications of this estimate in Section~\ref{sec:DM}.

In addition, de Boer et al. (2013) report the presence of two stellar overdensities in the Fornax dSph with lower luminosities than F6. These stellar systems also have elongated shapes (see their Fig. 2a), high metallicities ($\mathrm{[Fe/H]}\simeq -0.6$) and young stellar ages ($\sim 1.5\gyr$), similar to those of the Metal Rich component of the Fornax dSph. VST observations confirmed the presence of those substructures, and revealed two additional overdensities that share similar properties (Bate et al. 2015). More recently, using DES data Wang et al. (2020) find several high-density regions of high-metallicity \& young stellar ages, some of which were new and some of which were previously known (Coleman et al. 2004).
These substructures are therefore good candidates for being made of MR field stars captured by a population of massive dark objects. Spectroscopic measurements of their systemic velocity and proper motions, 
together with their internal velocity dispersion are needed in order to test this scenario.

The Eridanus II dSph is a dwarf spheroidal galaxy with low luminosity, $N_\star^{\rm dSph}\sim (6\pm 1)\times 10^4$ (Crnojevi{\'c} et al. 2016), a half-light radius of $r_h^{\rm dSph}=299\pm 12\pc$ (Simon et al. 2021), and low velocity dispersion $\sigma^{\rm dSph}=6.9\pm 1.1\kms$ (Li et al. 2017). Surprisingly for such a faint galaxy, it contains a single stellar cluster located at $R=23 \pm 3 \pc$ from its centre (Koposov et al. 2015). The cluster has very unusual properties, e.g. it is ancient $\sim 13.5\pm 0.3\gyr$, and extremely metal-poor $\mathrm{[Fe/H]}=-2.75\pm 0.2$ (Weisz et al. 2023). It has a remarkably large size, $r_h=15\pm 1\pc$, for a cluster with luminosity $M_V= -2.7\pm 0.3$ (Simon et al. 2021) (note that Crnojevi{\'c} et al. 2016 derive slightly different values). Like F6, Eri II cluster is elongated, $\epsilon\approx 0.31\pm 0.05$, with a shape that is remarkably well aligned with the ellipticity of the Eri II galaxy (Simon et al. 2021).
An elongated shape could be the telltale signature of these systems being close to full tidal destruction and therefore in a disequilibrium state. Yet, $N$-body models that adopt this scenario struggle to reproduce the extended size of these clusters, and predict bright tidal tails that should have already been observed in the data currently available (Orkney et al. 2022).
Crucially in the context of this paper, the age and metallicity of the Eri II cluster are statistically {\it indistinguishable} from the dSph itself (Crnojevi\'c et al. 2016; Simon et al. 2021; Weisz et al. 2023), which makes this system a good candidate for being composed of field stars captured by a dark substructure.
Again, spectroscopic follow-up measurements are needed to test the main prediction from this scenario, i.e whether the Eridanus II cluster exhibits anomalously high velocity dispersion compatible with being a DM-dominated system.

As a cautionary note, it must be pointed out that the nature of some of those stellar substructures is still debated. For example, there were early claims that F6 is not a stellar cluster, but actually an overdensity of unresolved background galaxies (Verner et al. 1981; Demers et al. 1994; Stetson et al. 1998). Only recently, DECam imaging and Gaia astrometric data have shown that F6 is clearly an overdensity of stars (Wang et al. 2019), a conclusion later supported by Magellan/M2FS spectroscopy, which helped to identify $\sim 15$--$17$ likely stellar members (Pace et al. 2021). 
Yet, the relatively small size of the data set may be still prone to systematics arising from low-number statistics. 

The unknown physical separation between these substructures and their host galaxies introduces additional uncertainty in the models. For simplicity, it is common to assume that these objects are located at a galactocentric distance equal to the projected separation on the sky, which is the minimum distance allowed by the data. Given that the phase-space density of field stars decreases with distance to the host galaxy centre, this choice can potentially bias future model constraints. For example, were these substructures located in the outskirts of the dwarf galaxies, substructures would need to have systematically larger masses in order to capture the same number of field stars (see \S\ref{sec:model}).

The results enclosed in this paper call for follow-up observational efforts to measure the spatial distribution, kinematics \& chemical composition of stellar substructures in dSphs with better accuracy, as they might provide unique constraints on the population of dark objects orbiting in these galaxies, as well as on the particle nature of dark matter, as briefly discussed in \S\ref{sec:DM}.

\subsection{Theory: simplifications and current uncertainties}\label{sec:theory}

The theoretical models presented in this work rely on a number of assumptions that are worth discussing here.

First, our substructure models source a spherical potential. 
Yet, anomalous stellar systems detected in dSphs, such as F6 and the lone cluster in Eridanus II, tend to exhibit elongated shapes that are well aligned with the morphology of the host galaxy (Wang et al. 2019; Simon et al. 2021). In the scenario proposed in this paper, stars trapped in a dark substructure
may appear aspherical due to an intrinsic triaxial shape of the objects that captured them, and/or a triaxial dwarf galaxy potential (e.g. Allgood et al. 2006; Despali et al. 2014). Interestingly, the numerical experiments plotted in Fig.~\ref{fig:xyz} show that spherical substructures accreted onto the galaxy on eccentric orbits may also generate flattend spatial overdensities that appear to be aligned with the orbital motion. 
Further theoretical work is needed to understand the possible connection the elongated shape of these stellar clusters and the expected triaxiality of dark matter haloes as well as their formation history.

The statistical experiments in \S\ref{sec:experiment} adopt substructure models with masses that are either constant, or grow with time. The latter models can be applied for example to study the population of field particles trapped around intermediate mass black holes (IMBHs) that grow in disks surrounding supermassive black holes (e.g. McKernan et al. 2012).
However, our models do not cover the case of self-gravitating subhaloes orbiting in a parent halo, which experience tidal stripping and mass loss after each pericentre passage (e.g. Pe\~narrubia et al. 2010). This shortcoming may be addressed by running live $N$-body simulations where substructures are modelled as self-gravitating objects. Unfortunately, the numerical tools required to explore this scenario are more complex and computationally expensive than the ones used for this work.
The results enclosed in this paper suggest that the analytical expressions outlined in \S\ref{sec:model} can be applied to time-dependent substructures, $\Phi_\bullet(t)$, insofar as its time-evolution is not impulsive, i.e.t it occurs on time scales $|\dot \Phi_\bullet/\Phi_\bullet|^{-1}\gtrsim T$, and the local approximation is reasonably accurate. Under these conditions, the population of bound field stars can be assumed to be in steady state in a time-varying potential $\Phi_\bullet(t)$, see Fig.~\ref{fig:numt_3}. For substructures moving on eccentric orbits around extended galaxy models, numerical experiments in \S\ref{sec:accretion}) show that the steady-state approximation may be reasonably accurate away from orbital pericentre. In truncated galaxy models, it is the local approximation that fails away from pericentre (see Fig.~\ref{fig:xyz}).

Our experiments do not explore a cosmologically-motivated scenario in which field stars form in a dark matter halo that contains a pre-existing population of dark matter clumps. By definition, dark substructures orbiting in these galaxies would not host a population of permanent captures, yet they may be able to trap field stars onto temporary orbits. Given the numerical experiments shown in \S\ref{sec:experiment}, we expect a different distribution of captured field stars compared to models where substructures grow in a pre-existing sea of particles. Cosmological hydrodynamical simulations that incorporate star formation/feedback are needed to tackle this issue.

The number, distribution and density profiles of dark matter sub-subhaloes in the satellite galaxies of MW-like haloes are notoriously uncertain.
For illustrative purposes, it is useful to estimate how many subhaloes that one would expect in a Fornax-like dwarf spheroidal with a virial mass of $M_{\rm vir}\sim 10^9\msol$ (e.g. Pe\~narrubia et al. 2008a; Errani et al. 2018). Adopting a mass ratio between the Fornax dSph and the MW of $\sim 10^9\msol /10^{12}\msol=10^{-3}$, and re-scaling CDM haloes of MW-like galaxies down three orders of magnitude suggests that the Fornax dSph should contain of the order of $\sim 100$ satellite sub-subhaloes with $M_\bullet\gtrsim 10^6\msol$ enclosed within its virial radius, $r^{\rm dSph}_{\rm vir}\sim 30\kpc$ (e.g. Weerasooriya et al. 2023 and references therein).

However, this naive scaling is only valid for field haloes. As pointed out above, dwarf spheroidals tend to be accreted early onto larger galaxies and lose a large fraction of their mass to Galactic tides (e.g. Pe\~narrubia \& Benson 2005; Errani et al. 2017), which possibly also removes a large fraction of the sub-subhalo population that fell in embedded with the dSph, particularly those with large orbital apocentres and long orbital periods (e.g Pe\~narrubia et al. 2008b; Errani \& Navarro 2021).

At present, it is not possible to turn to cosmological simulations of structure formation for accurate predictions on the number of dark sub-subhaloes in dSphs. Current $N$-body methods struggle to resolve objects with internal crossing times as short as $T\sim 1\myr$ in Milky Way-like haloes. Furthermore, these simulations suffer from well-known numerical artifacts (such as self-heating and artificial disruption) on scales comparable to the resolution of the simulation (e.g. van den Bosch \& Ogiya 2018; Errani \& Pe\~narrubia 2020). These issues call for dedicated high-resolution $N$-body simulations that address these numerical shortcomings and provide reliable statistics of ensembles of sub-subhaloes in dwarf spheroidal galaxies. 

Alternatively, since the orbits of MW dSphs are relatively well known thanks to the astrometry provided by the Gaia mission (e.g. Battaglia et al. 2022 and references therein), it is now feasible to model the properties of the surviving population of dark substructures in each individual MW dSph using constrained $N$-body simulations that do not suffer from the above-mentioned numerical artifacts. We leave these questions to future work.

\subsection{Dark Matter particle constraints}\label{sec:DM}
The detection and characterization of dark sub-subhaloes in dSphs may provide unprecedented constraints on the mass and self-interacting cross-section of DM particle models.
For example, the vast majority of gaps in cold tidal streams around MW-like galaxies are sensitive to subhaloes with masses in the range $10^6 < M_\bullet/\msol < 10^8$ (Erkal et al. 2016), while strong-lens observations are sensitive to subhaloes with masses $M_\bullet \gtrsim 6\times 10^8\msol$ (e.g. O'Riordan et al. 2023).
In contrast, Figs.~\ref{fig:mmin} and~\ref{fig:kml} show that dSphs may be able to probe subhalo masses down to $M_\bullet\sim 10^4$--$10^5\msol$. 

Measuring the masses of dark sub-subhaloes down to those scales could significantly improve current bounds on the free-streaming length of DM particles. This is best illustrated by quantifying the
half-mode mass $m_{hm}$, defined as the mass scale
where the WDM power
spectrum is suppressed by half with respect to CDM models (Schneider et al. 2012)  
$$m_{\rm hm}=3\times 10^8\msol \bigg(\frac{m_{\rm DM}}{3.3\kev}\bigg)^{-3.33}.$$
Detecting field stars trapped in a dark subhalo with a mass of $M_\bullet\simeq 10^6\msol \gtrsim m_{\rm hm}$ would imply a half-mass mode of $m_{\rm hm}\gtrsim 18\kev$, significantly tightening existing constraints from gravitational lensing ($m_{\rm DM}\gtrsim 5\kev$; Gilman et al. 2020), Lyman-alpha forest ($m_{\rm DM}\gtrsim 3\kev$; Villasenor et al. 2022), or the combination of strong gravitational lensing, the Ly-$\alpha$ forest, the number of luminous satellites in the Milky Way, which put a lower particle mass limit of $m_{\rm DM}\gtrsim 6\kev$ (Enzi et al. 2021; Nadler et al. 2021), and $m_{\rm DM}> 9.7$ keV when the Milky Way satellite population is combined with strong-lensing flux ratio statistics (Nadler et al. 2021). Detecting objects with masses $M_\bullet\simeq 10^5\msol \gtrsim m_{\rm hm}$ would push the constraints up to $m_{\rm hm}\gtrsim 37\kev$.

In addition, the internal structure of these systems may also reveal whether DM particles experience self-interactions. For example, dark substructures with $M_\bullet\sim 10^6 \msol$ must be sufficiently compact, $c_\bullet\lesssim 20\pc$, in order to generate a stellar overdensity of field stars in a Fornax-like dSph (see middle panel of Fig.~\ref{fig:kml}). This condition translates into a characteristic density $\rho_\bullet=M_\bullet/(2\pi c_\bullet^3)\gtrsim 20\msol\pc^{-3}$, which is higher than CDM subhaloes with similar masses. More precisely, field halos with a mass of $M_{200}=10^6\msol$ at redshift $z=0$ have a mean peak velocity radius of $r_{\rm max}\sim 100\pc$. Equating this radius to the substructure scale radius, $c_\bullet=r_{\rm max}$, yields a considerably lower characteristic density of $\rho_\bullet\sim 0.1\msol\pc^{-3}$ (Ludlow et al. 2016). 

On the one hand, the existence of DM sub-subhaloes with abnormally-high densities may point to interactions in the dark sector (e.g. Kahlhoefer et al. 2019). On the other, it is not immediately clear whether such a high density would necessarily be in conflict with CDM predictions.
First, because sub-subhalos in satellite galaxies are expected to be tidally processed, and more concentrated than field halos of the same mass (e.g. Pe\~narrubia et al. 2010; Errani \& Navarro 2021).
And second, because it is possible that the detection of dense sub-subhaloes may be due to observational biases. 
Indeed, compact substructures produce higher density enhancements in the field (see \S\ref{sec:model}), which can therefore be more easily detected as localized overdensities ($\delta_\star>1$). In this picture, anomalous objects like F6 may be sampling the high-density, low-probability tail of the sub-subhalo population. That is, they would represent the proverbial `tip of the iceberg', pointing to the presence of a much larger population of diffuse ($\delta_\star<1$) substructures with a power-law luminosity function~(\ref{eq:lumfunc}). There are several ways to test this prediction, e.g. by modelling statistical fluctuations of number counts in photometric surveys (e.g. Scheuer 1957), or by searching for clumps with low-velocity dispersion and/or significant velocity offsets in kinematic surveys of dSphs (e.g. Pace et al. 2014).

It is also worth noting that the presence of a large DM component in clusters such as F6 and the Eri II would have important implications for their survival in the host galaxy. 
For example, using collisional $N$-body simulations Contenta et al. (2019) show that the Eri II cluster (with no DM in the star cluster) quickly disrupt in a cuspy DM halo, favouring dSph models with a cored DM profile. 
In addition, Brandt (2016) shows that the survival of a stellar cluster near the centre of a dwarf galaxy depends on the number of massive compact halo objects populating the DM halo (see also Li et al. 2017; Zoutendijk et al. 2020), whereas Marsh \& Niemeyer (2019) point out that the cluster could also be used to test quantum fluctuations of ultra-light DM models. However, if these objects are embedded in dense DM haloes, they would become resilient to tidal stripping and would also be protected against collisions with nearby compact objects and rapid fluctuations of the local tidal field by dynamical invariance (e.g. Weinberg 1994; Pe\~narrubia 2019), weakening the constraints derived from their very survival.

\section{Summary}\label{sec:sum}

Substructures orbiting a larger galaxy can capture field stars that pass nearby with low relative velocities. In this work, stars become captured by the substructure potential $\Phi_\bullet$ when their gravitational energy $E=v^2/2+\Phi_\bullet(r)$ flips from positive to negative.
Our numerical experiments show that the orbits of stars captured through this mechanism can be separated into two families: (i) {\it Temporary captures}. The majority of field stars trapped in the potential $\Phi_\bullet$ move on chaotic orbits that orbit around the substructure for a finite amount of time before being released back to the galactic potential. This leads to a net number of temporary captures that converges to a steady state as the rate of capture becomes roughly equal to the escape rate. 
Stars captured on temporary orbits generate stellar overdensities ($\delta_\star>1)$ at the location of substructures that are sufficiently compact, $\kappa=1-c_\bullet/r_\epsilon>0$, where $r_\epsilon$ and $c_\bullet$ are the thermal critical radius~(\ref{eq:r_eps}) and scale radius of $\Phi_\bullet$, respectively.
(ii) {\it Permanent captures}. In addition, we find that immersing a
substructure potential in a galaxy that is in dynamical equilibrium at $t=0$ invariably results in a population of field particles that remain bound to $\Phi_\bullet$ for arbitrarily-long times. In static models where $\Phi_\bullet$ does not vary with time and moves on a circular orbit around the host galaxy, we find that particles on permanent orbits were already bound to the substructure potential at $t=0$, and remain within its tidal radius indefinitely. These particles are therefore reminiscent of the `potential escapers' discussed by Fukushige \& Heggie (2000).
Our experiments also show that substructures with a time-growing mass, $M_\bullet(t)$, can also capture stars on permanent orbits at $t>0$, with a rate that peaks at $t\sim \tau$, where $\tau=|\dot M_\bullet/M_\bullet|^{-1}_{t=0}$ is the time-scale that determines how fast the substructure mass varies. 
Importantly, we find that stars on permanent orbits have a density that lies below the host galaxy background ($\delta_\star<1$), and therefore do not generate visible overdensities.

The superposition of permanent and temporary orbits results in a population of captured field stars with an homogeneous phase-space distribution around the substructure potential, $Q_\star=n_\star/\sigma_\star^3\approx {\rm const}$. In steady state, these stars follow number density and dispersion profiles that scale as $n_\star\sim |\Phi_\bullet|^{3/2}$, and $\sigma_\star\sim |\Phi_\bullet|^{1/2}$, respectively (see \S\ref{sec:number}). Experimental results show that these theoretical predictions are accurate for substructures that source a static potential and move on a circular orbit across a sea of field particles initially in equilibrium. They become less accurate when applied to permanent \& temporary captures individually, or to substructures on very eccentric orbits moving through a rapidly-changing background.

In the current cosmological paradigm, galaxies are expected to host a large number of DM subhaloes, many of which may contain captured field stars. Here, we show that these objects would have a luminosity function governed by the subhalo mass function. 
Adopting a CDM-motivated mass function $\d N/\d M_\bullet\sim M_\bullet^{-\alpha}$ with $\alpha\simeq 1.9$ (e.g. Springel et al. 2008) leads to stellar substructures with a power-law luminosity function~(\ref{eq:lumfunc}), which scales as $\d N/\d M_\star\sim M_\star^{-\beta}$, where $\beta=(2\alpha+1)/3\approx 1.6$. However, only subhaloes that are sufficiently `compact' ($\kappa>0$) may materialize as localized stellar over-densities in the galactic background ($\delta_\star>1)$. 

In dwarf spheroidal galaxies, field stars captured by compact substructures may resemble stellar `clusters' with anomalous properties, e.g. they have extended sizes for their luminosity, contain stellar populations indistinguishable from the field, and exhibit DM-dominated mass-to-light ratios ($M/L\gg 1$). In contrast, field stars captured by `fluffly' substructures ($\kappa<0$) have densities below the background ($\delta_\star<1)$, and are therefore more difficult to detect. In spectroscopic surveys they may appear as clumps of co-moving field stars with low velocity dispersion ($\langle \sigma_\star^2\rangle^{1/2}<\sigma$) and/or a significant velocity offset ($V_\bullet \ne 0$). 
In cluster models that contain no DM, the presence of stellar systems with extended sizes and high mass-to-light ratios is typically attributed to on-going tidal disruption, which can be tested by searching for the associated tidal tails (Orkney et al. 2022). Yet, our experiments show that stars captured by substructures moving on eccentric orbits are released back to the galactic potential along tidal tails (see right panels Fig.~\ref{fig:xyz}), which complicates a clear-cut distinction between the two scenarios.

Capture models make a number of unique testable predictions. E.g. stellar overdensities composed of captured field stars have stellar ages and metallicities indistinguishable from those of the host galaxy, and high mass-to-light ratios $M/L\gg 1$ indicative of the presence of a substantial DM component. 
In galaxies with multiple chemo-dynamical populations, like the Fornax dSph, these models also make predictions on the ratio of captured stars as a function of metallicity. In addition, the density and velocity dispersion profiles of captured stars provide simultaneous constraints on the substructure mass profile, ($M_\bullet, c_\bullet)$, as well as on its systemic velocity with respect to the host galaxy, $V_\bullet$, which can be potentially measured with a combination of accurate astrometric \& spectroscopic data. The numerical experiments outlined in \S\ref{sec:experiment} show that the distribution of captured field stars also depends on the formation mechanism of the substructure, e.g. whether it grew within a pre-existing sea of field particles, or it formed outside of the galaxy and was accreted at a later time (see Fig.~\ref{fig:delta_trun}). This suggests that it may be possible to constrain the formation history, mass profile and systemic motion of the substructure simultaneously by fitting the distribution and kinematics of stellar systems with anomalous properties.

We have given numerical and theoretical arguments that indicate that `dark' subhaloes that were {\it not} sufficiently massive to trigger in-situ star formation may become `visible' by capturing baryonic particles
from the host galaxy field. The important implication of this result is that {\it dark subhaloes may not be completely invisible}. If they contain gravitationally-bound baryonic matter, they must emit and absorb radiation, which opens up a new avenue to test CDM predictions on halo mass scales that have not been probed to date. 
Follow-up work is needed to inspect the detectability of `dark' subhaloes in dSphs with masses below the star formation threshold, as well as observational campaigns aimed at detecting \& characterizing those objects.
In \S\ref{sec:DM} we discuss how the detection and characterization of dark subhaloes may provide unique \& unprecedented constraints on the particle mass and cross section for a large range of DM candidates.

\section*{Acknowledgements}
We are grateful to the referee for the constructive criticism and the suggestions that helped to improve the paper.
We thank Sergey Koposov, Alan McConnachie and Frank van den Bosch for their valuable comments and suggestions.
M.G.W. acknowledges support from National Science Foundation (NSF) grants AST-1813881, AST-1909584 and AST-2206046. M.G. acknowledges financial support from the grants PID2021-125485NB-C22, EUR2020-112157,  CEX2019-000918-M funded by MCIN/AEI/10.13039/501100011033 (State Agency for Research of the Spanish Ministry of Science and Innovation) and SGR-2021-01069 (AGAUR).
RE acknowledges support from the European Research Council (ERC) under the European Unions Horizon 2020 research and innovation programme (grant agreement number 834148), and from the National Science Foundation (NSF) grant AST-2206046. 
TB's research was supported by an appointment to the NASA Postdoctoral Program at the NASA Ames Research Center, administered by Oak Ridge Associated Universities under contract with NASA.

In memory of Teo Pe\~narrubia Reihl (16-3-2018, 6-1-2024). V.A.E.J.H.

\section*{Data Availability}
No data were generated for this study.
{}

\end{document}